\documentclass[journal]{IEEEtran}
\usepackage{multirow}
\usepackage{diagbox}
\usepackage{amssymb}
\usepackage{amsmath}
\usepackage{graphicx}
\usepackage{cite}
\usepackage{citesort}
\usepackage{subfigure}
\usepackage{graphicx,epstopdf}
\usepackage{epsfig}	
\usepackage{cite,graphicx,amsmath,amssymb}
\usepackage{comment}

\usepackage{amssymb}
\usepackage{amsmath}
\usepackage{cite}
\usepackage{url}
\usepackage{xcolor}
\usepackage{cite,graphicx,amsmath,amssymb}
\usepackage{subfigure}
\usepackage{citesort}
\usepackage{fancyhdr}
\usepackage{mdwmath}
\usepackage{mdwtab}
\usepackage{caption}
\usepackage{amsthm}
\usepackage{lipsum}

\newtheorem{theorem}{Theorem}

\newtheorem{lemma}{Lemma}

\newtheorem{corollary}{Corollary}

\newtheorem{remark}{Remark}  
\newtheorem{proposition}{Proposition}

\makeatletter
\def\ScaleIfNeeded{%
\ifdim\Gin@nat@width>\linewidth \linewidth \else \Gin@nat@width
\fi } \makeatother

\begin{document}

\title{Secure Communications in a Unified Non-Orthogonal Multiple Access Framework}

\author{ Xinwei~Yue,~\IEEEmembership{Member,~IEEE,} Yuanwei\ Liu,~\IEEEmembership{Member,~IEEE,} Yuanyuan~Yao,~\IEEEmembership{Member,~IEEE,} Xuehua Li, ~\IEEEmembership{Member,~IEEE,} Rongke Liu, ~\IEEEmembership{Member,~IEEE,} and Arumugam~Nallanathan,~\IEEEmembership{Fellow,~IEEE}

\thanks{X. Yue, Y. Yao and X. Li are with the School of Information and Communication Engineering and the Key Laboratory of Modern Measurement $\&$ Control Technology, Ministry of Education, Beijing Information Science and Technology University, Beijing 100101, China (email: \{xinwei.yue, yyyao and lixuehua\}@bistu.edu.cn).}
\thanks{Y. Liu and A. Nallanathan are with the School of Electronic Engineering and Computer Science, Queen Mary University of London, London E1 4NS, U.K. (email: \{yuanwei.liu, a.nallanathan\}@qmul.ac.uk).}
\thanks{R. Liu is with the School of Electronic and Information Engineering, Beihang University, Beijing 100191,
China (email: rongke$\_$liu@buaa.edu.cn). Part of this work has been submitted to IEEE ICC 2019 \cite{Yue2019PLSNOMA}.}
}


\maketitle

\begin{abstract}
This paper investigates the impact of physical layer secrecy on the performance of a unified non-orthogonal multiple access (NOMA) framework, where both external and internal eavesdropping scenarios are examined. The spatial locations of legitimate users (LUs) and eavesdroppers are modeled by invoking stochastic geometry. To characterize the security performance, new exact and asymptotic expressions of secrecy outage probability (SOP) are derived for both code-domain NOMA (CD-NOMA) and power-domain NOMA (PD-NOMA), in which imperfect successive interference cancellation (ipSIC) and perfect SIC (pSIC) are taken into account. For the external eavesdropping scenario, the secrecy diversity orders by a pair of LUs (the $n$-th user and $m$-th user) for CD/PD-NOMA are obtained. Analytical results make known that the diversity orders of the $n$-th user with ipSIC/pSIC for CD-NOMA and PD-NOMA are equal to zero/$K$ and zero/one, respectively. The diversity orders of the $m$-th user are equal to $K$/one for CD/PD-NOMA.
For the internal eavesdropping scenario, we examine the analysis of secrecy diversity order and observe that the $m$-th user to wiretap the $n$-th user with ipSIC/pSIC for CD-NOMA and PD-NOMA provide the diversity orders of zero/$K$ and zero/one, respectively, which is consistent with external eavesdropping scenario.
Numerical results are present to confirm the accuracy of the analytical results developed and show that: i) The secrecy outage behavior of the $n$-th user is superior to that of the $m$-th user; ii) By increasing the number of subcarriers,
CD-NOMA is capable of achieving a larger secrecy diversity gain compared to PD-NOMA.
\end{abstract}
\section{Introduction}
With the growing pervasiveness of smart devices and increasing the demand of mobile data traffic, it is necessary to explore a solution to spectrum resource shortage. The fifth generation (5G) mobile communication networks have aroused a great deal of concern, since it is capable of reflecting a large diversity of application domains and communication requirements \cite{Boccardi6736746,Wang6736752}.
Non-orthogonal multiple access (NOMA) has been envisaged as a promising technique of 5G networks to ensure massive connectivity and achieve higher spectrum utilization \cite{Ding2017Mag,Liu8114722}. The essential thought of NOMA is that a plurality of users share the same time-frequency resources over different power allocation with the aid of superposition coding. At receiver side, successive interference cancellation (SIC) or message passing algorithm (MPA) are carried out to extract the desired signals by mitigating multiuser interference \cite{Cover1991Elements,Tse2005}. From the perspective of standard trend, NOMA has been approved for inclusion in the 3rd generation partnership project long-term evolution advanced (LTE-A) systems \cite{MUST,Ding7973146Survey}, where NOMA is referred to as downlink multiuser superposition transmission scheme.
Furthermore, layered division multiplexing scheme has been accepted by the next generation digital TV standard (ATSC 3.0), which is one of non-orthogonal multiplexing techniques \cite{Zhang7378924TV}.

Many kinds of NOMA schemes have been proposed, which can be further divided into two major categories, i.e., power-domain NOMA (PD-NOMA) and code-domain NOMA (CD-NOMA). Until now, the academic mainly investigate the system performance of PD-NOMA and its applications. In \cite{Ding2014performance}, the performance of PD-NOMA has been characterized in terms of outage probability and ergodic rate with spatially random users. To guarantee the fairness of multiple users, the proportional fairness scheduling is employed to maximize sum rate of NOMA system \cite{Choi7527668Fairness}. As a further advance, the authors in \cite{Ding7273963Pairing} have studied the impact of user pairing on the performance of PD-NOMA, where the gap of sum rates between PD-NOMA and conventional multiple access is determined by the users' channel conditions. Applying NOMA into cooperative communications, the outage behaviors of cooperative PD-NOMA system have been evaluated experimentally in \cite{Ding2014Cooperative,Yue8026173Exploiting}, where the user with the better channel conditions is viewed as a decode-and-forward (DF) relay to deliver information. Triggered by this, the authors of \cite{Men7454773TVT,Yue7812773Fixed} have analyzed the outage performance of NOMA based on an amplify-and-forward (AF) relaying over Nakagami-$m$ fading scenarios. With the goal of maximizing energy efficiency, the application of simultaneous wireless information and power transfer to PD-NOMA has been discussed extensively in \cite{Liu7445146SWIPT}. Apart from the above studies, recently in \cite{Ding8368286Caching}, the authors have investigated the impact of PD-NOMA on mobile edge computing, where the offloading probability and energy consumption of user are evaluated respectively. To capture the potential performance of PD-NOMA in the context of wireless caching, the authors in
\cite{Kiani8267072Edge,Ding8467377EdgeComputing} have proposed a pair of PD-NOMA caching schemes ( i.e., push then delivery and push-and-delivery) and researched the caching hit probability and delivery outage probability.

Different from PD-NOMA, the main study of the industry belongs to CD-NOMA, which can be regarded as an extension of PD-NOMA. With regard to CD-NOMA, a large number of users' information are mapped into multiple resource elements or subcarriers by the virtue of a sparse spreading matrix. The remarkable representatives of CD-NOMA consist of sparse code multiple access \cite{Nikopour7037423SCMA}, pattern division multiple access \cite{Dai8352623PDMA}, multi-user sharing access \cite{Yuan2016MUSA} and resource spread multiple access, etc. From the viewpoint of a unified framework, the authors in \cite{Hanzo8352616} have identified various NOMA schemes, where the performance comparison of NOMA schemes is presented form receiver complexity, user overload and peak throughput.
In \cite{Qin8387207UDN}, resource allocation and user association were surveyed in the unified NOMA framework based on heterogeneous ultra dense networks. 
Exploiting the specific characteristics of the unified NOMA framework, the authors of \cite{Yue8370069Unified} have analyzed the outage performance and system throughput of CD/PD-NOMA, where the unified expressions of both exact and approximate expressions for connection outage probabilities are derived in detail.

Together with NOMA, physical layer security (PLS) has attracted more research attention. The issues of security features for 5G networks have taken on an increasingly pivotal role in the corresponding security services \cite{Mukherjee6739367,Fang8125684}.
Explicit insights for understanding NOMA and PLS, the maximization of secrecy sum rate for NOMA system was analyzed in \cite{Zhang7426798SecrecyRate}, where the power allocation to users is taken into consideration. In \cite{Ding7906532Multicast}, the authors have studied the impact of NOMA with multicasting transmission on PLS. By applying stochastic geometry theory, the authors of \cite{Liu7812773Security} have investigated the security outage behavior of single antenna and multiple antenna transmission scenarios for NOMA networks. Condition on unknown channel state information (CSI), the authors evaluated the secrecy outage behavior of NOMA system \cite{He7982788}, in which the design of decoding order and power allocation were formulated. To further enhance the secrecy performance of NOMA systems, the PLS of multiple-input single-output NOMA system was studied with the objective of protecting confidential information of legitimate users (LUs) by employing artificial noise \cite{Ding8309413}. Furthermore, the authors in \cite{Lei8333750SecrecyTAS} investigated the secrecy outage behaviors of multiple-input multiple-output NOMA, where the max-min transmit antenna selection scheme was proposed carefully.
Combining the PLS and cooperative communication, the secrecy rate of NOMA systems was analyzed in \cite{Arafa201801449} via introducing a number of trusted half-duplex relaying, i.e., cooperative jamming, DF and AF relaying. In the presence of external eavesdroppers (Eves), the authors of \cite{Zhang8335775} highlighted the security features of AF based two-way NOMA networks by exploiting subcarriers and user scheduling. In addition, the authors in \cite{Lei2018SecuringTrusted} investigated the secrecy outage performance of multiple-relays assisted NOMA with invoking the relay selection schemes.
\subsection{Motivations and Contributions}
While the aforementioned significant treatise has laid a basic foundation for understanding the applications of NOMA, the PLS of NOMA for a unified framework is far from being well understood. In \cite{Ding7906532Multicast}, the security enhancement of multicast-unicast scenario based on NOMA is achieved, where the internal eavesdropping scenario, i.e., the distant users are regarded as the Eves to wiretap the information of nearby users is only discussed. With the emphasis on PLS, the authors of \cite{Liu7812773Security} studied the secrecy performance of PD-NOMA, where the perfect successive interference cancellation (pSIC) scheme is carried out at LUs and Eves. This assumption will overestimate users' detection capability and result in the performance deviation. In addition, the authors of \cite{Yue8370069Unified} evaluated the outage behaviors of CD/PD-NOMA in a unified framework, but the secure communication scenario of unified NOMA framework is not surveyed.
Motivated by the aforementioned research contributions, we focus our attentions on investigating the PLS for a unified NOMA framework, where both external and internal eavesdropping scenarios are considered.
Relaxing the assumption of pSIC, imperfect successive interference cancellation (ipSIC) scheme is carried out to cover the implementation issues. Moreover, the spatial locations of LUs and Eves are modeled by using stochastic geometry.
The secrecy analyses for a pair of NOMA LUs are characterized in terms of secrecy outage probability (SOP).
Based on the above explanations, the primary contributions of this article are summarized as follows:
\begin{enumerate} 
  \item
   For external eavesdropping scenario, we derive the exact expressions of SOP for the pairing LUs (i.e., the $n$-th user and $m$-th user) in CD/PD-NOMA networks. We also derive the asymptotic SOP and obtain the corresponding secrecy diversity orders. We confirm that the secrecy diversity orders of the $n$-th user with pSIC for CD/PD-NOMA are equal to the number of subcarriers, i.e., $K$ and one, respectively. Due to the effect of residual interference, the SOP of the $n$-th user with ipSIC for CD/PD-NOMA converges to an error floor in the high signal-to-noise ratio (SNR) region and obtain a zero secrecy diversity order.
   \item
   We derive both exact and asymptotic expressions of SOP for the $m$-th user in CD/PD-NOMA networks. On the basis of asymptotic analyses, we identify that the secrecy diversity orders of the $m$-th user for CD/PD-NOMA are equal to $K$ and one, respectively. We observe that the secrecy outage performance of the $n$-th user with pSIC is superior to that of the $m$-th user. We also demonstrate that the secrecy outage behavior of the $n$-th user with pSIC is superior to that of the conventional multiple access (OMA) scheme, while the secrecy performance of the $m$-th user is inferior to OMA.
   \item
   For internal eavesdropping scenario, we derive the exact expressions of SOP for the $m$-th user to wiretap the information of the $n$-th user in CD/PD-NOMA networks. To obtain remarkable insights, we further derive the asymptotic SOP of this scenario and obtain the corresponding diversity order. On the condition of ipSIC, the secrecy diversity orders of the $m$-th user to wiretap the information of the $n$-th user for CD/PD-NOMA are equal to zero. We confirm that the corresponding SOP converges to an error floor in the high SNR region.
   \item
   We examine the secrecy diversity order of the $m$-th user to wiretap the information of the $n$-th user with pSIC.  We observe that the secrecy diversity orders for CD/PD-NOMA networks are equal to $K$ and one, respectively. We show that the security performance of CD-NOMA is better than that of PD-NOMA in internal eavesdropping scenario.
\end{enumerate}
\subsection{Organization and Notation}
The rest of this paper is organized as follows. Section \ref{Network Model} introduces the network model and transmission formulation of secure communications in a unified NOMA framework. In Section \ref{Connection Outage}, the secrecy performance of the unified NOMA framework are evaluated by using outage probability. Numerical results in Section \ref{Numerical Results} are shown for providing the available insights and performance evaluation. Finally,
Section \ref{Conclusion} summarizes this paper. The mathematical derivation proofs are collected in the appendix.

The main notations of this paper are shown as follows.
${F_X}\left(  \cdot  \right)$ and ${f_X}\left(  \cdot  \right)$ denote the cumulative distribution function (CDF) and probability density function (PDF) of a random variable $X$; The superscripts ${\left(  \cdot  \right)^*}$ and ${\left(  \cdot  \right)^T}$ stand for  conjugate-transpose operation and transpose, respectively; $\mathbb{E}\{\cdot\}$ denotes the expectation operation; $diag\left(  \cdot  \right)$ represents a diagonal matrix; $\left\|  \cdot  \right\|_2^2$ denotes Euclidean two norm of a vector; ${\bf{I}}_K$ is an $K\times K$ identity matrix.
\section{Network Model}\label{Network Model}
\begin{figure}[t!]
    \begin{center}
        \includegraphics[width= 3.1in, height=1.9in]{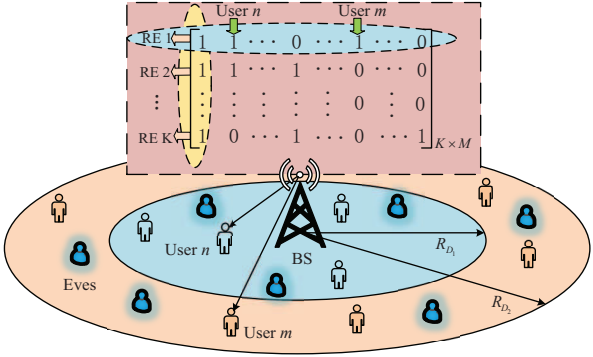}
         \caption{
         An illustration of a unified secure NOMA transmission network, where the spatial distributions of LUs and Eves follow homogeneous poisson point processes (HPPPs).}
        \label{PLS_NOMA_system_model}
    \end{center}
\end{figure}
\subsection{Network Descriptions}
Consider a unified secure NOMA transmission scenario illustrated in Fig. \ref{PLS_NOMA_system_model}, where a base station (BS) transmits the information to $M$ LUs in the presence of malicious Eves.
In particular, the BS maps the information of multiple LUs into $K$ subcarriers or time-frequency resource elements by utilizing a predefined sparse matrix ${{\bf{G}}_{K \times M}}$\footnote{The sparse matrix can be interpreted as the mapping from transmitted data information to subcarriers, which can be found in \cite{Nikopour7037423SCMA,Dai8352623PDMA}. A ``1'' means that the data shall be mapped to the corresponding subcarrier, while ``0'' means that no data is mapped.}, where the matrix dimension satisfies the inequality $1 \le K < M$. The aim of this relationship is to ensure the overload of sparse matrix. If $M$ is less than or equal to $K$, it will make sure each user to enjoy an orthogonal resource element, while in the non-orthogonal case, $M$ is greater than $K$. There are a few number of non-zero entries within ${{\bf{G}}_{K \times M}}$ and each non-zero entry corresponds to a subcarrier.
To facilitate analysis, the BS, LUs and Eves are equipped with single antenna, respectively. As a further advance, assuming that $M$ LUs are divided into $M/2$ orthogonal pairs, where the nearby user and distant user are capable of being distinguished based on their disparate channel conditions. The BS is located at the center of a disc denoted as $D_1$, with radius $R_{D_{1}}$ and $M/2$ users are distributed within $D_1$, while the remaining $M/2$ users are deployed within the disc denoted as $D_2$, with radius $R_{D_{2}}$. The spatial locations of LUs and Eves are governed by marked HPPPs ${\Phi _l}$ and ${\Phi _e}$ with density ${\lambda _l}$ and ${\lambda _e}$ \cite{chiu2013stochastic,Liu7982794Heterogeneous}, respectively. A bounded path loss model is used to model the channel coefficients from the BS to LUs and Eves, which is to ensure that the path loss is always larger than one for any distance. The channel between BS and LUs, and between BS and Eves as are denoted as main channel and wiretap channel, respectively.
Here we emphasize that the CSI of Eves is not accessible at the BS while that of LUs is available. A pair of users are randomly selected to carry out NOMA protocol\footnote{The benefit of user pairing is that co-channel interference and system complexity are relatively small, which will lead to the better SOP performance for PD/CD-NOMA. Since multiple users are admitted at the same time/frequency and spreading code, co-channel interference and system complexity will be strong in NOMA systems \cite{Ding7273963Pairing,Lei8006199Selection}.}. It is worth pointing out that more sophisticated user pairing is capable of enhancing the system performance, which will be researched in our future works. From the perspective of secure transmission design, we consider both external and internal eavesdropping scenarios in this treatise. In external eavesdropping scenario, the Eves individually overhear the communication of main channel between the BS and LUs, where the $n$-th user and $m$-th user are randomly selected to carry out the NOMA protocol; In internal eavesdropping scenario, the $m$-th user who has the worse channel conditions is viewed as Eve to wiretap the $n$-th user. We assume that the nearby users and Eves are not simultaneously distributed on the same boundary, but considering the case of the nearby users and Eves appearing on the same boundary is capable of further enriching the contents, which is beyond the scope of this paper. 
\subsection{Signal Model}
In unified NOMA transmission scenario,
the data streams of each user, i.e., $M$ LUs are spread by employing one column of sparse matrix and then superposed over $K$ subcarriers. Hence when the BS transmits the superposed signals to LUs, the observation at the $\varphi $-th user for the unified NOMA framework is given by
\begin{align}\label{The received signal expression at the n and m-th user}
{{\bf{y}}_\varphi }{\rm{ = }}diag\left( {{{\bf{h}}_\varphi }} \right) ( {{{\bf{g}}_n}\sqrt {{P_s}{a_n}} {x_n} + {{\bf{g}}_m}\sqrt {{P_s}{a_m}} {x_m}} ) + {{\bf{n}}_\varphi },
\end{align}
where $\varphi  \in \left( {n,m} \right)$. $x_{n}$ and $x_{m}$ are supposed to be normalized power signals of the $n$-th user and $m$-th user, respectively, i.e, $\mathbb{E}\{x_{n}^2\}= \mathbb{E}\{x_{m}^2\}=1$. Additionally, we assume that the power allocation coefficients of the $n$-th user and $m$-th user satisfy the conditions ${a_m} > {a_n}$ and $a_m + a_n = 1$, which is taken into account for fairness between LUs.
The optimal power allocation coefficients are capable of enhancing the performance of NOMA networks, which are beyond the scope of this paper. $P_{s}$ denotes the BS's transmission power. 
${{\bf{g}}_\varphi } = {[{g_{\varphi 1}}{g_{\varphi 2}} \cdots {g_{\varphi K}}]^T}$ is the mapping indicator vector of the $\varphi $-th user which is one column of ${{\bf{G}}_{K \times M}}$. More specifically, ${g_{\varphi k}}=1$ and ${g_{\varphi k}}=0$ indicate whether or not the signals are mapped into the $k$-th subcarrier, respectively. For the sake of clarity, assuming that ${{\bf{G}}_{K \times M}}$ belongs to regular sparse matrix with the same column weight. ${{\bf{h}}_\varphi } = {[{{\tilde h}_{\varphi 1}}{{\tilde h}_{\varphi 2}} \cdots {{\tilde h}_{\varphi K}}]^T}$ denotes the channel vector between the BS and $\varphi $-th user at $K$ subcarriers with ${{\tilde h}_{\varphi k}} = \frac{{\sqrt \eta  {h_{\varphi k}}}}{{\sqrt {1 + {d_\varphi ^\alpha }} }}$, where ${h_{\varphi k}} \sim {\cal C}{\cal N}\left( {0,1} \right)$ is the Rayleigh fading channel between the BS and $\varphi $-th user at the $k$-th subcarrier,
$\eta$ is a frequency dependent factor, $\alpha $ is the path loss exponent and ${d_\varphi }$ is the distance from BS to the $\varphi $-th user.
${{\bf{n}}_\varphi } \sim {\cal C}{\cal N}\left( {0,{N_0}{{\bf{I}}_{K}}} \right)$ denotes the additive white Gaussian noise (AWGN) at the $\varphi $-th user.
Note that based on the setting of the number of subcarriers, the unified NOMA framework can be categorized to CD-NOMA ($ K \ne 1$) and PD-NOMA ($K=1$), respectively. From the viewpoint of complexity order, the complex of CD-NOMA is greater than that of PD-NOMA\footnote{Since the operation of multiplications is much more complex than that of additions, many contributions \cite{Tse2005,Hu4444758} compare the complexity with respect to the number of multiplications. At this time, the complexity of MPA for CD-NOMA is on the order of $O( {{d_f}K{M^{{d_f}}}} )$, where $M$ and ${d_f}$ denote the size of modulation constellation and maximum row weight of sparse matrix, respectively, while the complexity order of SIC is equal to one for PD-NOMA.}.

To maximize the output SNRs, the maximal-ratio combining (MRC) is carried out \cite{Tse2005,Yang6327665} at the $\varphi$-th user over $K$ subcarriers. Let ${{\mathbf{u}}_\varphi } = \frac{{{{\left( {diag\left( {{{\mathbf{h}}_\varphi }} \right){{\mathbf{g}}_\varphi }} \right)}^*}}}{{\left\| {diag\left( {{{\mathbf{h}}_\varphi }}\right){{\mathbf{g}}_\varphi }} \right\|}}$ and the observation at the $\varphi $-th user can be rewritten as
\begin{align}\label{The further received signal expression at the n and m-th user}
{{\tilde y}_\varphi }{\text{ = }}{{\mathbf{u}}_\varphi }diag\left( {{{\mathbf{h}}_\varphi }} \right) ( {{{\mathbf{g}}_n}\sqrt {{P_s}{a_n}} {x_n} + {{\mathbf{g}}_m}\sqrt {{P_s}{a_m}} {x_m}} ) + {{\mathbf{u}}_\varphi }{{\mathbf{n}}_\varphi }.
\end{align}
Due to having better channel condition, the $n$-th user will carry out the SIC scheme, i.e.,
the $n$-th user first detect the information $x_m$ and remove this component from the superposed signal to decode the information $x_n$. Hence the signal-plus-interference-to-noise ratio (SINR) of the $n$-th user to detect its own information is give by
\begin{align}\label{the SINR expression at n-th LU to detect itself with SIC}
{\gamma _n} = \frac{{\rho \left\| {diag\left( {{{\bf{h}}_n}} \right){{\bf{g}}_n}} \right\|_2^2{a_n}}}{{\varpi \rho \left\| {{{\bf{h}}_I}} \right\|_2^2 + 1}},
\end{align}
where $\rho  = \frac{{{P_s}}}{{{N_0}}}$ denotes the transmit SNR. ${\varpi} \in \left[ {0,1} \right]$ is the impact level of residual interference. i.e., $\varpi {\rm{ \ne  0}}$ and $\varpi {\rm{ = 0}}$ denote the ipSIC and pSIC operation, respectively.
${{\bf{h}}_I} = {[{h_{I1}}{h_{I2}} \cdots {h_{IK}}]^T}$ denotes the residual interference channel vector with $ {h_{Ik}} \sim {\cal C}{\cal N}\left( {0,{\Omega _I}} \right)$.

The SINR of the $m$-th user to decode the information of itself is given by
\begin{align}\label{SINR m}
{\gamma _m} = \frac{{\rho \left\| {diag\left( {{{\bf{h}}_m}} \right){{\bf{g}}_m}} \right\|_2^2{a_m}}}{{\rho \left\| {diag\left( {{{\bf{h}}_m}} \right){{\bf{g}}_n}} \right\|_2^2{a_n} + 1}}.
\end{align}
\subsubsection{External Eavesdropping Scenario}
Different from \cite{Liu7812773Security}, we focus on relaxing the idealistic assumptions that the Eves are incapable of carrying out the perfect multiuser detection \footnote{It is worth mentioning that assuming the Eves have not enough powerful detection abilities is more in line with the requirements of practical scenarios, which is also the similar assumption in \cite{Lei8006199Selection}.}. For instance, the Eves only have limited computing power and allow the interference signal to exist. Similar to \eqref{The further received signal expression at the n and m-th user}, the MRC-combined signal at Eves for the unified framework can be given by
\begin{align}\label{The further received signal expression at Eve}
{{\tilde y}_e }{\text{ = }}{{\mathbf{u}}_e }diag\left( {{{\mathbf{h}}_e }} \right) ( {{{\mathbf{g}}_n}\sqrt {{P_s}{a_n}} {x_n} + {{\mathbf{g}}_m}\sqrt {{P_s}{a_m}} {x_m}} ) + {{\mathbf{u}}_e }{{\mathbf{n}}_e },
\end{align}
where ${{\mathbf{u}}_e } = \frac{{{{\left( {diag\left( {{{\mathbf{h}}_e }} \right){{\mathbf{g}}_e }} \right)}^*}}}{{\left\| {diag\left( {{{\mathbf{h}}_e }}\right){{\mathbf{g}}_e }} \right\|}}$. ${{\bf{h}}_e} = {[{{\tilde h}_{e1}}{{\tilde h}_{e2}} \cdots {{\tilde h}_{eK}}]^T}$ denotes the channel vector between the BS and Eves over $K$ subcarriers with ${{\tilde h}_{ek}} = \frac{{\sqrt \eta  {h_{ek}}}}{{\sqrt {1 + {d_{e}^\alpha }} }}$, where ${h_{ek}} \sim {\cal C}{\cal N}\left( {0,1} \right)$ is Rayleigh fading channel from the BS to Eves at the $k$-th subcarrier, and $d_{e}$ is the distance between BS and Eves. ${{\bf{g}}_e} = {[{g_{e1}}{g_{e2}} \cdots {g_{eK}}]^T}$ belongs to one column of ${{\bf{G}}_{K \times M}}$. ${{\bf{n}}_e } \sim {\cal C}{\cal N}\left( {0,{N_e}{{\bf{I}}_{K}}} \right)$ denotes the AWGN at Eves.

Based on the above assumptions, the Eves first detect the $m$-th user's information and regard the $n$-th user's information as interference signal. Then the received SINRs of detecting the $n$-th user's information at the most pernicious Eve can be given by
\begin{align}\label{the SINR expression of eavesdropper to detect the n-th user}
{\gamma _{{E_n}}} = \mathop {\max }\limits_{e \in {\Phi _e}} \left\{ {\frac{{{\rho _e}\left\| {diag\left( {{{\bf{h}}_e}} \right){{\bf{g}}_e}} \right\|_2^2 {a_n}}}{{\varpi {\rho _e}\left\| {{{\bf{h}}_{{I_e}}}} \right\|_2^2 + 1}}} \right\},
\end{align}
where ${\rho _e} = \frac{{{P_s}}}{{{N_e}}}$ denote the transmit SNR with variance ${N_e}$ of AWGN at Eves and
${{\bf{h}}_{{I_e}}} = {[{h_{{I_e}1}}{h_{{I_e}2}} \cdots {h_{{I_e}K}}]^T}$ denotes the residual interference channel vector with ${h_{{I_e}k}} \sim {\cal C}{\cal N}\left( {0,{\Omega _{{I_e}}}} \right)$.

The SINR of detecting the $m$-th user's information at the most pernicious Eve is given by
\begin{align}\label{the SINR expression of eavesdropper to detect the m-th user}
{\gamma _{{E_m}}} = \mathop {\max }\limits_{e \in {\Phi _e}} \left\{ {\frac{{{a_m}\left\| {diag\left( {{{\bf{h}}_e}} \right){{\bf{g}}_e}} \right\|_2^2}{\rho _e}}{{{a_n}\left\| {diag\left( {{{\bf{h}}_e}} \right){{\bf{g}}_e}} \right\|_2^2 {\rho _e} + 1}}} \right\}.
\end{align}
\subsubsection{Internal Eavesdropping Scenario}
The distant user (i.e., the $m$-th user) is regarded as Eve, who directly wiretap the nearby user's (the $n$-th user) information. At this moment, the SNR of the $m$-th user to detect the $n$-th user can be given by
\begin{align}\label{the SINR expression of eavesdropper to the nearby user}
{\gamma _{{E_{{m \to n}}}}} = \mathop {\max }\limits_{e \in {\Phi _e}} \left\{ {\rho_e \left\| {diag\left( {{{\bf{h}}_m}} \right){{\bf{g}}_m}} \right\|_2^2 {a_n}} \right\}.
\end{align}
\subsection{Channel Statistical Formulations}
In this subsection, we present the channel statistical properties of main channel, which will be used to evaluate SOP in the following section.
\begin{lemma}\label{Lemmad: The_CDF of near LUs with ipSIC for CD-NOMA}
Assuming $M$ LUs distributed uniformly within a disc $D_1$, the CDF $F_{CD,{\gamma _n}}^{ipSIC}$ of the $n$-th user with ipSIC for CD-NOMA is given by
\begin{align}\label{CD-NOMA:the CDF of SINR expression for the near LUs with ipSIC}
F_{CD,{\gamma _n}}^{ipSIC}\left( x \right) & \approx \Delta \sum\limits_{u = 1}^U {{b_u}\left[ {\Omega _I^K\Gamma \left( K \right) - {e^{ - \frac{{x{c_u}}}{{\eta \rho {a_n}}}}}} \right.} \sum\limits_{i = 0}^{K - 1} {\sum\limits_{j = 0}^i {\frac{{\phi \psi }}{{i!}}} } \nonumber \\
  &\times {\left( {\frac{{x{c_u}}}{{\eta \rho {a_n}}}} \right)^i}\left. {{{\left( {\frac{{\eta {a_n}{\Omega _I}}}{{x\varpi {c_u}{\Omega _I} + \eta {a_n}}}} \right)}^{K + j}}} \right],
\end{align}
where $\Delta  = \frac{1}{{\left( {K - 1} \right){\rm{!}}\Omega _I^K}}$, ${b_u} = \frac{\pi }{{2U}}\sqrt {1 - \theta _u^2} \left( {{\theta _u} + 1} \right)$, ${c_u}{\rm{ = }}1 + {\left[ {\frac{{{R_{D_1}}}}{2}\left( {{\theta _u} + 1} \right)} \right]^\alpha }$, ${\theta _u} = \cos \left( {\frac{{2u - 1}}{{2U}}\pi } \right)$, $\phi  = {
   i  \choose
   j  } = \frac{{i{\rm{!}}}}{{\left( {i - j} \right){\rm{!}}j{\rm{!}}}}$, $\psi  = {\left( {\varpi \rho } \right)^j}\Gamma \left( {K + j} \right)$, $\varpi {\rm{ \ne  0}}$ and $U$ is a parameter to ensure a complexity-accuracy tradeoff. $\Gamma \left(  \cdot  \right)$ denotes the Gamma function \cite[Eq. (8.310.1)]{gradshteyn}. 
 \begin{proof}
See Appendix~A.
\end{proof}
\end{lemma}
Upon substituting $\varpi {\text{ = }}0$ into \eqref{CD-NOMA:the CDF of SINR expression for the near LUs with ipSIC}, the CDF $F_{CD,{\gamma _n}}^{pSIC}$  of the $n$-th user with pSIC for CD-NOMA can be given by
\begin{align}\label{the CDF of gamma_n with pSIC for CD-NOMA}
F_{CD,{\gamma _n}}^{pSIC}\left( x \right) \approx \sum\limits_{u = 1}^U {{b_u}\left[ {1 - {e^{ - \frac{{x{c_u}}}{{\eta \rho {a_n}}}}}\sum\limits_{i = 0}^{K - 1} {\frac{1}{{i!}}{{\left( {\frac{{x{c_u}}}{{\eta \rho {a_n}}}} \right)}^i}} } \right]} .
\end{align}

For the special case with $K=1$, the CDF $F_{PD,{\gamma _n}}^{ipSIC}$ of the $n$-th user with ipSIC for PD-NOMA is given by
\begin{align}\label{PD-NOMA:the CDF of SINR expression for the near LUs with ipSIC}
F_{PD,{\gamma _n}}^{ipSIC}\left( x \right) \approx \sum\limits_{u = 1}^U {{b_u}} \left( {1 - \frac{{\eta {a_n}}}{{\eta {a_n} + x{c_u}{\Omega _I}}}{e^{ - \frac{{x{c_u}}}{{\eta \rho {a_n}}}}}} \right).
\end{align}

Upon substituting $\varpi {\text{ = }}0$ into \eqref{PD-NOMA:the CDF of SINR expression for the near LUs with ipSIC}, the CDF $F_{PD,{\gamma _n}}^{ipSIC}$ of the $n$-th user with pSIC for PD-NOMA is given by
\begin{align}\label{the CDF of gamma_n with pSIC for PD-NOMA}
F_{PD,{\gamma _n}}^{pSIC}\left( x \right) \approx \sum\limits_{u = 1}^U {{b_u}} \left( {1 - {e^{ - \frac{{x{c_u}}}{{\eta \rho {a_n}}}}}} \right).
\end{align}
\section{Secrecy Performance Analysis}\label{Connection Outage}
In this section, we focus attention on the scenario, where the CSI of the Eves is not available at BS. More precisely, the BS transmits the information at a fixed secrecy data rate. Under this condition, if instantaneous secrecy rate is less than a target secrecy rate, secrecy outage will be happen\cite{Gopala4626059,Ekrem5730586}. Hence the SOP\footnote{It is worth noting that SOP is a well-known metric to evaluate secrecy protocols. However, the practical suitability of this criterion is beyond of this treatise.} can be employed to characterize the security performance for the unified NOMA framework.
Inspired by this, the security performance of both external and internal eavesdropping scenarios are characterized in terms of secrecy outage behaviors. To this end, both exact and asymptotic expressions of SOP for LUs are provided in detail. Based on the analytical results, secrecy diversity order of the unified NOMA network can be obtained in the high SNR region.  Note that more meaningful comparisons (i.e., secrecy ergodic rate and energy efficiency) between PD-NOMA and CD-NOMA will further enrich the contents of this paper, which we may include in our future work.

\subsection{External Eavesdropping Scenario}
Considering a two user case, the $n$-th user and $m$-th user are paired together to carry out the NOMA protocol. Define the secrecy rate achieved of the $n$-th user and $m$-th user for the unified NOMA framework as
\begin{align}\label{the secrecy rate for near user}
{C_n} = {\left[ {{{\log }_2}\left( {1 + {\gamma _{_n}}} \right) - {{\log }_2}\left( {1 + {\gamma _{{E_n}}}} \right)} \right]^ + },
\end{align}
and
\begin{align}\label{the secrecy rate for far user}
{C_m} = {\left[ {{{\log }_2}\left( {1 + {\gamma _{_m}}} \right) - {{\log }_2}\left( {1 + {\gamma _{{E_m}}}} \right)} \right]^ + },
\end{align}
respectively, where ${\left( x \right)^ + } = \max \left\{ {0,x} \right\}$.
\subsubsection{The SOP of the $n$-th user}
Based on the above definitions, when the secrecy rate of the $n$-th user is less than the target secrecy rate, the secrecy outage will occur. Hence the SOP of the $n$-th user in external eavesdropping scenario can be expressed as
\begin{align}\label{CD-NOMA:the expression of COP for near user}
P_{n}\left( {{R_n}} \right) =& {\rm{Pr}}\left( {{C_n} < {R_n}} \right)    \nonumber\\
=& \int_0^\infty  {f_{{\gamma _{{E_n}}}}\left( x \right)} F_{{\gamma _n}}\left( {{2^{{R_n}}}\left( {1 + x} \right) - 1} \right)dx,
\end{align}
where ${R_n}$ denotes the $n$-th user's target secrecy rate.

The follow-up focus is that the PDF ${f_{{\gamma _{{E_n}}}}}$ of wiretap channel should be formulated accurately. Furthermore, the following lemma provides the PDF ${f_{CD,{\gamma _{{E_n}}}}^{ipSIC}}$ of Eves with ipSIC for CD-NOMA in detail.
\begin{lemma}\label{CD-NOMA:the_PDF for near user with ipSIC}
Assuming the Eves distributed uniformly within the disc $D_2$, the PDF ${f_{CD,{\gamma _{{E_n}}}}^{ipSIC}}$ of the most pernicious Eve with ipSIC for CD-NOMA is given by \eqref{CD-NOMA:the PDF expression for the Eves with ipSIC} at the top of next page, where $\Phi {\rm{ = }} - 2\pi {\lambda _e}$, ${\varphi _1}{\rm{ = }}\eta {\rho _e}{a_n}$, ${\varphi _2}{\rm{ = }}{\rho _e}{\Omega _{{I_e}}}$,
$\zeta  = \frac{{{{\left( {\varpi {\varphi _2}} \right)}^j}\left( {K + j - 1} \right)!}}{{\varphi _1^{ - K - j + i}\left( {K - 1} \right)!}}$ and $\varpi {\rm{ \ne  0}}$.
\begin{proof}
See Appendix~B. 
\end{proof}
\begin{figure*}[!t]
\normalsize
\begin{align}\label{CD-NOMA:the PDF expression for the Eves with ipSIC}
f_{CD,{\gamma _{{E_n}}}}^{ipSIC}\left( x \right) =& {e^{\Phi \sum\limits_{i = 0}^{K - 1} {\sum\limits_{j = 0}^i {\frac{{{x^i}\phi \zeta }}{{i!}}} } \int_0^\infty  {{{\left[ {{\varphi _1} + x\varpi {\varphi _2}\left( {1 + {r^\alpha }} \right)} \right]}^{ - K - j}}{{\left( {1 + {r^\alpha }} \right)}^i}{e^{ - \frac{{x\left( {1 + {r^\alpha }} \right)}}{{{\varphi _1}}}}}rdr} }}\Phi \sum\limits_{i = 0}^{K - 1} {\sum\limits_{j = 0}^i {\frac{{\phi \zeta }}{{i!}}} } {e^{ - \frac{x}{{{\varphi _1}}}}} \nonumber \\
  &\times \int_0^\infty  {\left\{ {\frac{{i{x^{i - 1}} - {x^i}\left( {1 + {r^\alpha }} \right)\varphi _1^{ - 1}}}{{{{\left[ {{\varphi _1} + x\varpi {\varphi _2}\left( {1 + {r^\alpha }} \right)} \right]}^{K + j}}}} - \frac{{\varpi {\varphi _2}\left( {1 + {r^\alpha }} \right)\left( {K + j} \right){x^i}}}{{{{\left[ {{\varphi _1} + x\varpi {\varphi _2}\left( {1 + {r^\alpha }} \right)} \right]}^{K + j + 1}}}}} \right\}} {e^{ - \frac{{x{r^\alpha }}}{{{\varphi _1}}}}}{\left( {1 + {r^\alpha }} \right)^i}rdr .
\end{align}
\hrulefill \vspace*{0pt}
\end{figure*}
\end{lemma}

Similar to the solution process in \eqref{CD-NOMA:the PDF expression for the Eves with ipSIC}, the PDF ${f_{CD,{\gamma _{{E_n}}}}^{pSIC}}$ of the most pernicious Eve with pSIC ($\varpi  = 0$) for CD-NOMA can be given by \eqref{CD-NOMA:the PDF expression of the Eves with pSIC} at the top of this page, where $\delta  = \frac{2}{\alpha }$.
\begin{figure*}[!t]
\normalsize
\begin{align}\label{CD-NOMA:the PDF expression of the Eves with pSIC}
f_{CD,{\gamma _{{E_n}}}}^{pSIC}\left( x \right) = & {e^{ - \delta \pi {\lambda _e}\sum\limits_{i = 0}^{K - 1} {\sum\limits_{j = 0}^i {{
   i  \choose
   j
}\frac{1}{{i!}}\varphi _1^{j - i + \delta  + 1}\Gamma \left( {j + \delta } \right){x^{i - 1 - j - \delta }}{e^{\frac{{ - x}}{{{\varphi _1}}}}}} } }} \nonumber \\
  &\times \delta \pi {\lambda _e}\sum\limits_{i = 0}^{K - 1} {\sum\limits_{j = 0}^i {{
   i  \choose
   j
}\frac{1}{{i!}}\varphi _1^{j - i + \delta  + 1}\Gamma \left( {j + \delta } \right)} } \left[ {\left( {j - i + 1 + \delta } \right){x^{i - j - \delta  - 2}} + \frac{{{x^{i - j - 1 - \delta }}}}{{{\varphi _1}}}} \right]{e^{\frac{{ - x}}{{{\varphi _1}}}}}.
\end{align}
\hrulefill \vspace*{0pt}
\end{figure*}

Upon substituting \eqref{CD-NOMA:the CDF of SINR expression for the near LUs with ipSIC} and \eqref{CD-NOMA:the PDF expression for the Eves with ipSIC} into \eqref{CD-NOMA:the expression of COP for near user}, the following theorem provides the exact analysis of SOP for the $n$-th user with ipSIC in CD-NOMA.
\begin{theorem}\label{Theorem:CD-NOMA:the COP of near user}
Conditioned on the HPPPs, the SOP of the $n$-th user with ipSIC for CD-NOMA is given by \eqref{the expression for the n-th user with ipSIC for CD-NOMA} at the top of next page, where $\vartheta = \frac{{{c_u}\left( {{2^{{R_n}}}\left( {1 + x} \right) - 1} \right)}}{{\eta \rho {a_n}}}$.
\begin{figure*}[!t]
\normalsize
\begin{align}\label{the expression for the n-th user with ipSIC for CD-NOMA}
P_{CD,n}^{ipSIC}\left( {{R_n}} \right) = & \int_0^\infty  {{e^{\Phi \sum\limits_{i = 0}^{K - 1} {\sum\limits_{j = 0}^i {\frac{{{x^i}\phi \zeta }}{{i!}}} } \int_0^\infty  {{{\left[ {{\varphi _1} + x\varpi {\varphi _2}\left( {1 + {r^\alpha }} \right)} \right]}^{ - K - j}}{{\left( {1 + {r^\alpha }} \right)}^i}{e^{ - \frac{{x\left( {1 + {r^\alpha }} \right)}}{{{\varphi _1}}}}}rdr} }}\Phi \sum\limits_{i = 0}^{K - 1} {\sum\limits_{j = 0}^i {\frac{{\phi \zeta }}{{i!}}} } {e^{ - \frac{x}{{{\varphi _1}}}}}} \nonumber \\
 &\times \int_0^\infty  {\left\{ {\frac{{i{x^{i - 1}} - {x^i}\left( {1 + {r^\alpha }} \right)\varphi _1^{ - 1}}}{{{{\left[ {{\varphi _1} + x\varpi {\varphi _2}\left( {1 + {r^\alpha }} \right)} \right]}^{K + j}}}} - \frac{{\varpi {\varphi _2}\left( {1 + {r^\alpha }} \right)\left( {K + j} \right){x^i}}}{{{{\left[ {{\varphi _1} + x\varpi {\varphi _2}\left( {1 + {r^\alpha }} \right)} \right]}^{K + j + 1}}}}} \right\}} {e^{ - \frac{{x{r^\alpha }}}{{{\varphi _1}}}}}{\left( {1 + {r^\alpha }} \right)^i}rdr  \nonumber \\
  & \times \Delta \sum\limits_{u = 1}^U {{b_u}} \left\{ {\Omega _I^K\Gamma \left( K \right) - {e^{ - \vartheta }}\sum\limits_{i = 0}^{K - 1} {\sum\limits_{j = 0}^i {\frac{{\phi \psi {\vartheta ^i}}}{{i!}}} } {{\left[ {\frac{{{\Omega _I}}}{{\varpi \rho \vartheta {\Omega _I} + 1}}} \right]}^{K + j}}} \right\}dx.
\end{align}
\hrulefill \vspace*{0pt}
\end{figure*}
\end{theorem}

As a further development, substituting \eqref{the CDF of gamma_n with pSIC for CD-NOMA} and \eqref{CD-NOMA:the PDF expression of the Eves with pSIC} into \eqref{CD-NOMA:the expression of COP for near user}, the SOP of the $n$-th user with pSIC for CD-NOMA can be given by \eqref{CD-NOMA:the expression for the m-th user with pSIC} at the top of next page.
\begin{figure*}[!t]
\normalsize
\begin{align}\label{CD-NOMA:the expression for the m-th user with pSIC}
P_{CD,n}^{pSIC}\left( {{R_n}} \right) = & \int_0^\infty  {{e^{ - \delta \pi {\lambda _e}\sum\limits_{i = 0}^{K - 1} {\sum\limits_{j = 0}^i {{
   i  \choose
   j
}\frac{1}{{i!}}\varphi _1^{j - i + \delta  + 1}\Gamma \left( {j + \delta } \right){x^{i - 1 - j - \delta }}{e^{\frac{{ - x}}{{{\varphi _1}}}}}} } }}} \delta \pi {\lambda _e}\sum\limits_{i = 0}^{K - 1} {\sum\limits_{j = 0}^i {{
   i  \choose
   j
}\frac{1}{{i!}}\varphi _1^{j - i + \delta  + 1}} } \nonumber \\
 & \times \Gamma \left( {j + \delta } \right) {e^{\frac{{ - x}}{{{\varphi _1}}}}}  \left[ {\left( {j - i + 1 + \delta } \right){x^{i - j - \delta  - 2}} + \frac{{{x^{i - j - 1 - \delta }}}}{{{\varphi _1}}}} \right]\sum\limits_{u = 1}^U {{b_u}} \left[ {1 - {e^{ - \vartheta }}\sum\limits_{i = 0}^{K - 1} {\frac{{{\vartheta ^i}}}{{i!}}} } \right]dx.
\end{align}
\hrulefill \vspace*{0pt}
\end{figure*}
\begin{corollary}\label{Corollary:PD-NOMA:the COP of far user}
For the special case with $K=1$, the SOP of the $n$-th user with ipSIC for PD-NOMA is given by \eqref{the expression for the n-th user with ipSIC for PD-NOMA}, where $\Theta  = - \delta \pi {\lambda _e}{\varphi _1}\Gamma \left( \delta  \right){e^{\varphi _2^{ - 1}}}$ and $\Gamma \left( { \cdot , \cdot } \right)$ denotes the upper incomplete Gamma function.
\begin{figure*}[!t]
\normalsize
\begin{align}\label{the expression for the n-th user with ipSIC for PD-NOMA}
P_{PD,n}^{ipSIC}\left( {{R_n}} \right) =& \Theta \int_0^\infty  {{e^{\frac{{\Theta {{\left( {{\varphi _1} + x{\varphi _2}} \right)}^{\delta  - 1}}}}{{{{\left( {x{\varphi _2}} \right)}^\delta }}}\Gamma \left( {1 - \delta ,\frac{{{\varphi _1} + x{\varphi _2}}}{{{\varphi _1}{\varphi _2}}}} \right)}}} \sum\limits_{u = 1}^U {{b_u}} \left( {1 - \frac{{{e^{ - \vartheta }}}}{{1 + \vartheta \rho {\Omega _I}}}} \right)\left\{ {\Gamma \left( {1 - \delta ,\frac{{{\varphi _1} + x{\varphi _2}}}{{{\varphi _1}{\varphi _2}}}} \right)} \right. \nonumber \\
&  \times \left. {\left[ {\frac{{\left( {\delta  - 1} \right){{\left( {{\varphi _1} + x{\varphi _2}} \right)}^{\delta  - 2}}}}{{{x^\delta }\varphi _2^{\delta  - 1}}} - \frac{{\delta {{\left( {{\varphi _1} + x{\rho _e}{\Omega _{{I_e}}}} \right)}^{\delta  - 1}}}}{{{x^{\delta  + 1}}\varphi _2^\delta }}} \right] - \frac{{\varphi _1^{\delta  - 1}{e^{ - \frac{{{\varphi _1} + x{\varphi _2}}}{{{\varphi _1}{\varphi _2}}}}}}}{{{x^\delta }\left( {{\varphi _1} + x{\varphi _2}} \right)}}} \right\}dx.
\end{align}
\hrulefill \vspace*{0pt}
\end{figure*}
\begin{proof}
See Appendix~C. 
\end{proof}
\end{corollary}

On the basis of derived process in Appendix C, substituting $\varpi {\rm{ = 0}}$ into \eqref{the expression for the n-th user with ipSIC for PD-NOMA}, the SOP of the $n$-th user with pSIC for PD-NOMA can be given by
\begin{align}\label{the SOP of the n-th user with pSIC for PD-NOMA}
P_{PD,n}^{pSIC}\left( {{R_n}} \right) =  &\sum\limits_{u = 1}^U {\frac{{\mu {b_u}}}{{\eta {\rho _e}{a_n}}}} \int_0^\infty  {\left( {\frac{1}{{{x^\delta }}} + \frac{{\eta {\rho _e}{a_n}\delta }}{{{x^{\delta  + 1}}}}} \right)}  \nonumber \\
 & \times {e^{ - \frac{\mu }{{{x^\delta }}}{e^{ - \frac{x}{{\eta {\rho _e}{a_n}}}}} - \frac{x}{{\eta {\rho _e}{a_n}}}}}\left( {1 - {e^{ - \vartheta }}} \right)dx,
\end{align}
where $\mu  = \delta \pi {\lambda _e}{\left( {\eta {\rho _e}{a_n}} \right)^\delta }\Gamma \left( \delta  \right)$.

\subsubsection{The SOP of the $m$-th user}
When the secrecy rate of the $m$-th user is less than target secrecy rate, secrecy outage will be declared. As such, the SOP of the $m$-th user can be expressed as
\begin{align}\label{the expression of SOP for the m-th user}
 P_{m}\left( {{R_m}} \right) = &{\rm{Pr}}\left( {{C_m} < {R_m}} \right) \nonumber\\
  = & \int_0^\infty  {{f_{{\gamma _{{E_m}}}}}\left( x \right)} {F_{{\gamma _m}}}\left( {{2^{{R_m}}}\left( {1 + x} \right) - 1} \right)dx,
\end{align}
where $R_m$ denotes the $m$-th user's target secrecy rate.


The exact analysis of SOP for the $m$-th user can be obtained in the following theorem.
\begin{theorem}\label{Theorem:the SOP of the m-th user for CD-NOMA}
Conditioned on the HPPPs, the SOP of the $m$-th user for CD-NOMA is given by \eqref{the SOP of the m-th user for CD-NOMA} at the next page, where $\tau  = \frac{1}{{{2^{{R_m}}}\left( {1 - {a_m}} \right)}} - 1$.
\begin{figure*}[!t]
\normalsize
\begin{align}\label{the SOP of the m-th user for CD-NOMA}
P_m^{CD}\left( {{R_m}} \right) =&\int_0^\tau  {{e^{ - \delta \pi {\lambda _e}\sum\limits_{i = 0}^{K - 1} {\sum\limits_{j = 0}^i {{
   i  \choose
   j
}\frac{{\Gamma \left( {j + \delta } \right)}}{{i!}}} } {{\left[ {\frac{{\eta {\rho _e}\left( {{a_m} - {a_n}x} \right)}}{x}} \right]}^{j + \delta  - i}}{e^{\frac{{ - x}}{{\eta {\rho _e}\left( {{a_m} - {a_n}x} \right)}}}}}}\delta \pi {\lambda _e}\sum\limits_{i = 0}^{K - 1} {\sum\limits_{j = 0}^i {{
   i  \choose
   j
}} \frac{{\Gamma \left( {j + \delta } \right)}}{{i!}}} {{\left( {\eta {\rho _e}} \right)}^{j + \delta  - i}}}  \nonumber \\
  &\times {e^{\frac{{ - x}}{{\eta {\rho _e}\left( {{a_m} - {a_n}x} \right)}}}}{\left( {{a_m} - {a_n}x} \right)^{j + \delta  - i}}\left[ {\frac{{{a_n}\left( {j - i + \delta } \right)}}{{{x^{j - i + \delta }}\left( {{a_m} - {a_n}x} \right)}} + \frac{{{a_m}}}{{\eta {\rho _e}{x^{j - i + \delta }}{{\left( {{a_m} - {a_n}x} \right)}^2}}} + \frac{{\left( {j - i + \delta } \right)}}{{{x^{j - i + \delta {\rm{ + }}1}}}}} \right] \nonumber \\
  &\times \sum\limits_{u = 1}^U {{b_u}} \left\{ {1 - {e^{ - \frac{{{c_u}\left( {{2^{{R_m}}}\left( {1 + x} \right) - 1} \right)}}{{\eta \rho \left( {{a_m} - \left( {{2^{{R_m}}}\left( {1 + x} \right) - 1} \right){a_n}} \right)}}}}\sum\limits_{i = 0}^{K - 1} {\frac{1}{{i!}}{{\left[ {\frac{{{c_u}\left( {{2^{{R_m}}}\left( {1 + x} \right) - 1} \right)}}{{\eta \rho \left( {{a_m} - \left( {{2^{{R_m}}}\left( {1 + x} \right) - 1} \right){a_n}} \right)}}} \right]}^i}} } \right\}dx .
\end{align}
\hrulefill \vspace*{0pt}
\end{figure*}
\begin{proof}
See Appendix~D.
\end{proof}
\end{theorem}
For the special case with $K=1$, the SOP of the $m$-th user for PD-NOMA is given by
\begin{align}\label{the SOP of the m-th user for PD-NOMA}
&P_m^{PD}\left( {{R_m}} \right) = \kappa \sum\limits_{u = 1}^U {{b_u}} \int_0^{{\tau }} {{e^{ - \frac{{\kappa {{\left( {{a_m} - {a_n}x} \right)}^\delta }}}{{{x^\delta }}}{e^{ - \frac{x}{{\eta {\rho _e}\left( {{a_m} - {a_n}x} \right)}}}}}}}  \nonumber\\
 & \times {e^{ - \frac{x}{{\eta {\rho _e}\left( {{a_m} - {a_n}x} \right)}}}}\left[ {\frac{{{a_n}\delta {{\left( {{a_m} - {a_n}x} \right)}^{\delta  - 1}}}}{{{x^\delta }}} + \frac{{\delta {{\left( {{a_m} - {a_n}x} \right)}^\delta }}}{{{x^{\delta  + 1}}}}} \right. \nonumber \\
& \left. { + \frac{{{a_m}{{\left( {{a_m} - {a_n}x} \right)}^{\delta  - 2}}}}{{\eta {\rho _e}{x^\delta }}}} \right]\left[ {1 - {e^{ - \frac{{\left( {{2^{{R_m}}}\left( {1 + x} \right) - 1} \right){c_n}}}{{\eta \rho \left( {{a_m} - \left( {{2^{{R_m}}}\left( {1 + x} \right) - 1} \right){a_n}} \right)}}}}} \right]dx,
\end{align}
where $\kappa  = \delta \pi {\lambda _e}{\left( {\eta {\rho _e}} \right)^\delta }\Gamma \left( \delta  \right)$.

\begin{proposition}\label{proposition:The SOP of the selected user pair}
The SOP of user pairing, i.e., the $n$-th user and $m$-th user with ipSIC/pSIC for CD/PD-NOMA are given by
\begin{align}\label{the SOP of the selected user pair for CD-NOMA}
P_{CD,nm}^\xi  = 1 - \left( {1 - P_{CD,n}^\xi } \right)\left( {1 - P_{CD,m}^\xi } \right),
\end{align}
and
\begin{align}\label{the SOP of the selected user pair for PD-NOMA}
P_{PD,nm}^\xi  = 1 - \left( {1 - P_{PD,n}^\xi } \right)\left( {1 - P_{PD,m}^\xi } \right),
\end{align}
respectively, where $\xi  \in \left( {ipSIC,pSIC} \right)$. $P_{CD,n}^{ipSIC}$, $P_{CD,n}^{pSIC}$ and $P_{CD,m}$ can be obtained from \eqref{the expression for the n-th user with ipSIC for CD-NOMA}, \eqref{CD-NOMA:the expression for the m-th user with pSIC} and \eqref{the SOP of the m-th user for CD-NOMA}, respectively. $P_{PD,n}^{ipSIC}$, $P_{PD,n}^{pSIC}$ and $P_{PD,m}$ can be obtained from \eqref{the expression for the n-th user with ipSIC for PD-NOMA}, \eqref{the SOP of the n-th user with pSIC for PD-NOMA} and \eqref{the SOP of the m-th user for PD-NOMA}, respectively.
\end{proposition}
\subsection{Internal Eavesdropping Scenario}
Assuming that the $m$-th user is viewed as Eve to wiretap the information of the $n$-th user. Hence the secrecy rate achieved by internal eavesdropping scenario can be expressed as
\begin{align}\label{the secrecy rate for far user to detect near user }
{C_{m \to n}} = {\left[ {{{\log }_2}\left( {1 + {\gamma _{_n}}} \right) - {{\log }_2}\left( {1 + {\gamma _{{E_{m \to n}}}}} \right)} \right]^ + }.
\end{align}

By using the above the definitions, when the achievable secrecy rate is inferior to the target secrecy rate, the SOP will be happen. Hence the SOP of the $m$-th user to wiretap the information of the $n$-th user can be given by
\begin{align}\label{the secrecy OP for far user to detect near user }
& P_{m \to n}^{CD}\left( {{R_{m \to n}}} \right)= {\rm{Pr}}\left( {{C_{m \to n}} < {R_{m \to n}}} \right) \nonumber \\
  &= \int_0^\infty  {{f_{{\gamma _{{E_{m \to n}}}}}}\left( x \right)} {F_{{\gamma _{n}}}}\left( {{2^{{R_{m \to n}}}}\left( {1 + x} \right) - 1} \right)dx,
\end{align}
where ${R_{m \to n}}$ denotes the target secrecy rate of the $m$-th user to wiretap the $n$-th user.

Similar to solution process of \eqref{CD-NOMA:the PDF expression for the Eves with ipSIC} for external eavesdropping scenario, the exact analysis of PDF $f_{{\gamma _{{E_{m \to n}}}}}^{CD}$ for CD-NOMA in internal eavesdropping scenario can be given by
\begin{align}\label{the PDF of far user for CD-NOMA in IE case}
&f_{{\gamma _{{E_{m \to n}}}}}^{CD}\left( x \right) = {e^{ - \Lambda \sum\limits_{i = 0}^{K - 1} {\sum\limits_{j = 0}^i {{
   i  \choose
   j
}\frac{{\Gamma \left( {j + \delta } \right)}}{{i!}}{{\left( {\frac{{\eta {\rho _e}{a_n}}}{x}} \right)}^{j + \delta  - i + 1}}{e^{\frac{{ - x}}{{\eta {\rho _e}{a_n}}}}}} } }} \nonumber \\
 &  \times \Lambda {e^{\frac{{ - x}}{{\eta \rho_{e} {a_n}}}}}\sum\limits_{i = 0}^{K - 1} {\sum\limits_{j = 0}^i {{
   i  \choose
   j
}\frac{1}{{i!}}\Gamma \left( {j + \delta } \right){{\left( {\eta \rho_{e} {a_n}} \right)}^{j - i + \delta  + 1}}} }   \nonumber\\
 & \times \left[ {\left( {{j + \delta  - i + 1}} \right){x^{i - j - \delta  - 2}} + \frac{{{x^{i - j - 1 - \delta }}}}{{\eta \rho_{e} {a_n}}}} \right] ,
\end{align}
where $ \Lambda = \delta \pi {\lambda _e}$.

In internal eavesdropping scenario, the CDF of ${\gamma _n}$ for the $n$-th user in the main channel is not be affected effectively. As a consequence, substituting \eqref{CD-NOMA:the CDF of SINR expression for the near LUs with ipSIC} and \eqref{the PDF of far user for CD-NOMA in IE case} into \eqref{the secrecy OP for far user to detect near user }, the following theorem is capable of providing the exact expression of SOP for the $m$-th user to wiretap the $n$-th user.

\begin{theorem}\label{Theorem:CD-NOMA:the COP of the m-th user for IE case}
Conditioned on the HPPPs, the SOP of the $m$-th user to wiretap the $n$-th user with ipSIC for CD-NOMA is given by \eqref{the SOP of the m-th user to wiretap the n-th user with ipSIC for CD-NOMA} at the top of next page, where ${\vartheta _1} = \frac{{{c_u}\left( {{2^{{R_{m \to n}}}}\left( {1 + x} \right) - 1} \right)}}{{\eta \rho {a_n}}}$.
\begin{figure*}[!t]
\normalsize
\begin{align}\label{the SOP of the m-th user to wiretap the n-th user with ipSIC for CD-NOMA}
&P_{m \to n}^{CD,ipSIC}\left( {{R_{m \to n}}} \right) = \Delta \Lambda \int_0^\infty  {{e^{ - \Lambda \sum\limits_{i = 0}^{K - 1} {\sum\limits_{j = 0}^i {{
   i  \choose
   j  }\frac{{\Gamma \left( {j + \delta } \right)}}{{i!}}{{\left( {\frac{{\eta {\rho _e}{a_n}}}{x}} \right)}^{j + \delta  - i + 1}}{e^{\frac{{ - x}}{{\eta {\rho _e}{a_n}}}}}} } }}} \sum\limits_{i = 0}^{K - 1} {\sum\limits_{j = 0}^i {{
   i  \choose
   j  }\frac{{\Gamma \left( {j + \delta } \right)}}{{i!}}} } {e^{\frac{{ - x}}{{\eta {\rho _e}{a_n}}}}}{\left( {\eta {\rho _e}{a_n}} \right)^{j + \delta  - i + 1}} \nonumber \\
 & \times \left[ {\left( {j + \delta  - i + 1} \right){x^{i - j - \delta  - 2}} + \frac{{{x^{i - j - 1 - \delta }}}}{{\eta {\rho _e}{a_n}}}} \right]\sum\limits_{u = 1}^U {{b_u}} \left\{ {\Omega _I^K\Gamma \left( K \right) - {e^{ - {\vartheta _1}}}\sum\limits_{i = 0}^{K - 1} {\sum\limits_{j = 0}^i {\frac{{\phi \psi {\vartheta _1}^i}}{{i!}}} } {{\left[ {\frac{1}{{\varpi \rho {\vartheta _1}{\Omega _I} + 1}}} \right]}^{K + j}}} \right\}dx .
\end{align}
\hrulefill \vspace*{0pt}
\end{figure*}
\end{theorem}

Upon substituting $\varpi  = 0$ into \eqref{the SOP of the m-th user to wiretap the n-th user with ipSIC for CD-NOMA} and after some manipulations, the SOP of the $m$-th user to wiretap the $n$-th user with pSIC for CD-NOMA is given by \eqref{the SOP of the m-th user to wiretap the n-th user with pSIC for CD-NOMA} at the top of next page.
\begin{figure*}[!t]
\normalsize
\begin{align}\label{the SOP of the m-th user to wiretap the n-th user with pSIC for CD-NOMA}
P_{m \to n}^{CD,pSIC}\left( {{R_{m \to n}}} \right) =& \Lambda \int_0^\infty  {{e^{ - \Lambda \sum\limits_{i = 0}^{K - 1} {\sum\limits_{j = 0}^i {{
   i  \choose
   j  }\frac{{\Gamma \left( {j + \delta } \right)}}{{i!}}{{\left( {\frac{{\eta {\rho _e}{a_n}}}{x}} \right)}^{j + \delta  - i + 1}}{e^{\frac{{ - x}}{{\eta {\rho _e}{a_n}}}}}} } }}}  \sum\limits_{i = 0}^{K - 1} {\sum\limits_{j = 0}^i {{
   i  \choose
   j  }\frac{{\Gamma \left( {j + \delta } \right)}}{{i!}}} } {e^{\frac{{ - x}}{{\eta {\rho _e}{a_n}}}}} \nonumber \\
 & \times {\left( {\eta {\rho _e}{a_n}} \right)^{j + \delta  - i + 1}}  \left[ {\left( {{j + \delta  - i + 1}} \right){x^{i - j - \delta  - 2}} + \frac{{{x^{i - j - 1 - \delta }}}}{{\eta {\rho _e}{a_n}}}} \right]\sum\limits_{u = 1}^U {{b_u}\left[ {1 - {e^{ - {\vartheta _1}}}\sum\limits_{i = 0}^{K - 1} {\frac{{\vartheta _1^i}}{{i!}}} } \right]}  dx .
\end{align}
\hrulefill \vspace*{0pt}
\end{figure*}

For the special case with $K=1$, the SOP of the $m$-th user to wiretap the $n$-th user with ipSIC for PD-NOMA is given by
\begin{align}\label{the SOP of the m-th user to wiretap the n-th user with ipSIC for PD-NOMA}
&P_{m \to n}^{PD,ipSIC}\left( {{R_{m \to n}}} \right) = \int_0^\infty  {\mu {e^{ - \frac{\mu }{{{x^\delta }}}{e^{ - \frac{x}{{\eta \rho {a_n}}}}} - \frac{x}{{\eta \rho {a_n}}}}}}  \nonumber \\
 & \times \left( {\frac{1}{{\eta \rho {a_n}{x^\delta }}} + \frac{\delta }{{{x^{\delta {\rm{ + }}1}}}}} \right)\sum\limits_{u = 1}^U {{b_u}} \left[ {1 - \frac{{{e^{ - {\vartheta _1}}}}}{{1 + \rho {\vartheta _1}{\Omega _I}}}} \right]dx  .
\end{align}

Upon substituting $\varpi  = 0$ into \eqref{the SOP of the m-th user to wiretap the n-th user with ipSIC for PD-NOMA}, the SOP of the $m$-th user to wiretap the $n$-th user with pSIC for PD-NOMA is given by
\begin{align}\label{the SOP of the m-th user to wiretap the n-th user with pSIC for PD-NOMA}
&P_{m \to n}^{PD,pSIC}\left( {{R_{m \to n}}} \right) = \int_0^\infty  {\mu {e^{ - \mu \frac{1}{{{x^\delta }}}{e^{ - \frac{x}{{\eta \rho_{e} {a_n}}}}} - \frac{x}{{\eta \rho_{e} {a_n}}}}}}  \nonumber \\
&  \times \left\{ {\frac{1}{{\eta \rho_{e} {a_n}{x^\delta }}} + \frac{\delta }{{{x^{\delta {\rm{ + }}1}}}}} \right\}\sum\limits_{u = 1}^U {{b_u}\left( {1 - {e^{ - {\vartheta _1}}}} \right)}.
\end{align}
\subsection{Secrecy Diversity Order Analysis}
In order to obtain more insights, the secrecy diversity order is selected to evaluate the secrecy performance of the unified NOMA framework, which is capable of describing how fast the SOP decreases with average SNR. Moreover, we proceed to derive the asymptotic SOP in the high SNR region, when the SNR of main channel between BS and LUs tends to infinity, i.e., $\rho  \to \infty $.
It is worth noting that $\rho  \to \infty $ corresponds to the scenario where the LUs is closed to BS than Eves, which is a practical scenario of interest. On the contrary, when the average SNR approaches infinity at Eves, the Eves can always wiretap information of LUs successfully. Hence the secrecy diversity order can be written as follows:
\begin{align}\label{The definition of diversity order for SOP}
d =  - \mathop {\lim }\limits_{\rho  \to \infty } \frac{{\log \left( {P_\infty \left( \rho  \right)} \right)}}{{\log \rho }},
\end{align}
where ${P_\infty \left( \rho  \right)}$ denotes the asymptotic SOP in the high SNR region.
\subsubsection{External Eavesdropping Scenario}
In light of the above explanations, when $\rho \to \infty $, we provide the approximate SOP of LUs for CD/PD-NOMA as follows.
\begin{corollary}\label{corollary:The asymptotic CDF of  the n-th user with ipSIC for CD-NOMA}
The asymptotic SOP of the $n$-th user with ipSIC at high SNR region for CD-NOMA is given by
\begin{align}\label{The asymptotic SOP of the n-th user with ipSIC for CD-NOMA}
&P_{CD,n,\infty }^{ipSIC}\left( {{R_n}} \right) = \Delta \int_0^\infty  {f_{CD,{\gamma _{{E_n}}}}^{ipSIC}} \left( x \right)\sum\limits_{u = 1}^U {{b_u}} \left\{ {\Omega _I^K\Gamma \left( K \right)} \right. \nonumber \\
 &  - \sum\limits_{i = 0}^{K - 1} {\frac{{{{\left( {\varpi \rho \vartheta } \right)}^i}}}{{i!}}} \left. {{{\left( {\frac{{{\Omega _I}}}{{\varpi \rho \vartheta {\Omega _I} + 1}}} \right)}^{K + i}}\Gamma \left( {K + i} \right)} \right\}dx  ,
\end{align}
where $\varpi {\rm{ \ne  0}}$ and ${f_{CD,{\gamma _{{E_n}}}}^{ipSIC}}$ can be obtained from \eqref{CD-NOMA:the PDF expression for the Eves with ipSIC}.
\begin{proof}
We show concern for diversity order analysis by characterizing the CDF of the LUs. When $\rho \to \infty $,
based on \eqref{Appenxix AA: The CDF of n-th user with ipSIC for CD-NOMA} and through some manuscripts, we can obtain the asymptotic unordered CDF of $F_{CD,{\gamma _n}}^{ipSIC}$ as follows:
\begin{align}\label{The asymptotic CDF of the n-th user with ipSIC for CD-NOMA}
& F_{CD,{\gamma _n}}^{ipSIC,\infty }\left( x \right) = \Delta \sum\limits_{u = 1}^U {{b_u}} \left[ {\Omega _I^K\Gamma \left( K \right) - \sum\limits_{i = 0}^{K - 1} {\frac{1}{{i!}}} } \right. \nonumber \\
 & \left. { \times {{\left( {\frac{{x\varpi {c_u}}}{{\eta {a_n}}}} \right)}^i}{{\left( {\frac{{\eta {a_n}{\Omega _I}}}{{x\varpi {c_u}{\Omega _I} + \eta {a_n}}}} \right)}^{K + i}}\Gamma \left( {K + i} \right)} \right] .
\end{align}
Then based on \eqref{CD-NOMA:the expression of COP for near user}, we can replace the CDF of $F_{CD,{\gamma _n}}^{ipSIC}$ by the asymptotic CDF $F_{CD,{\gamma _n}}^{ipSIC,\infty }$ and arrive at the asymptotic SOP of the $n$-th user with ipSIC for CD-NOMA. The proof is completed.
\end{proof}
\end{corollary}

\begin{remark}\label{remark1: the n-th user with ipSIC for CD-NOMA}
Up on substituting \eqref{The asymptotic SOP of the n-th user with ipSIC for CD-NOMA} into \eqref{The definition of diversity order for SOP}, we observe that the secrecy diversity order of the $n$-th user with ipSIC for CD-NOMA is equal to zero. This is due to the fact that there is the effect of residual interference from ipSIC.
\end{remark}

On the basis of the approximate result in \cite[Eq. (29)]{Yue8370069Unified} and removing the ordering operation,
the asymptotic SOP of the $n$-th user with pSIC for CD-NOMA at high SNR region is given by
\begin{align}\label{The asymptotic SOP of the n-th user with pSIC for CD-NOMA}
P_{CD,n,\infty }^{pSIC}\left( {{R_n}} \right) = \sum\limits_{u = 1}^U {\frac{{{b_u}}}{{K!}}} \int_0^\infty  {f_{CD,{\gamma _{{E_n}}}}^{pSIC}\left( x \right)} {\vartheta ^K}dx,
\end{align}
where ${f_{CD,{\gamma _{{E_n}}}}^{pSIC}}$ can be obtained from \eqref{CD-NOMA:the PDF expression of the Eves with pSIC}.
\begin{remark}\label{remark2: the n-th user with pSIC for CD-NOMA}
Up on substituting \eqref{The asymptotic SOP of the n-th user with pSIC for CD-NOMA} into \eqref{The definition of diversity order for SOP}, the secrecy diversity order of the $n$-th user with pSIC for CD-NOMA is equal to the number of subcarriers $K$. It is shown that CD-NOMA is capable of achieving more secrecy diversity gain compared to PD-NOMA.
\end{remark}

Similar to \eqref{The asymptotic SOP of the n-th user with ipSIC for CD-NOMA} and \eqref{The asymptotic SOP of the n-th user with pSIC for CD-NOMA}, the asymptotic SOP of the $n$-th user with ipSIC/pSIC at high SNR region for PD-NOMA are given by
\begin{align}\label{The asymptotic SOP of the n-th user with ipSIC for PD-NOMA}
&P_{PD,n,\infty }^{ipSIC}\left( {{R_n}} \right) = \sum\limits_{u = 1}^U {{b_u}} \int_0^\infty  {f_{PD,{E_n}}^{ipSIC}\left( x \right)} \left( {\frac{{\vartheta \rho {\Omega _I}}}{{1 + \vartheta \rho {\Omega _I}}}} \right)dx ,
\end{align}
and
\begin{align}\label{The asymptotic SOP of the n-th user with pSIC for PD-NOMA}
P_{PD,n,\infty }^{pSIC}\left( {{R_n}} \right) = \sum\limits_{u = 1}^U {{b_u}} \int_0^\infty  {f_{PD,{E_n}}^{pSIC}\left( x \right)\vartheta } dx,
\end{align}
respectively, where $f_{PD,{E_n}}^{ipSIC}\left( x \right) = \Theta {e^{\frac{{\Theta G{{\left( {{\varphi _1} + x{\varphi _2}} \right)}^{\delta  - 1}}}}{{{{\left( {x{\varphi _2}} \right)}^\delta }}}}} \\
\times \left( {QG - \frac{{\varphi _1^{\delta  - 1}{e^{ - \frac{{{\varphi _1} + x{\varphi _2}}}{{{\varphi _1}{\varphi _2}}}}}}}{{{x^\delta }\left( {{\varphi _1} + x{\varphi _2}} \right)}}} \right)$, $Q = \frac{{\left( {\delta  - 1} \right){{\left( {{\varphi _1} + x{\varphi _2}} \right)}^{\delta  - 2}}}}{{{x^\delta }\varphi _2^{\delta  - 1}}} - \frac{{\delta {{\left( {{\varphi _1} + x{\rho _e}{\Omega _{{I_e}}}} \right)}^{\delta  - 1}}}}{{{x^{\delta  + 1}}\varphi _2^\delta }}$, $G = \Gamma \left( {1 - \delta ,\frac{{{\varphi _1} + x{\varphi _2}}}{{{\varphi _1}{\varphi _2}}}} \right)$.
$f_{PD,{\gamma _{{E_n}}}}^{pSIC}\left( x \right) = \mu \Gamma \left( \delta  \right){e^{ - \mu \Gamma \left( \delta  \right)\frac{1}{{{x^\delta }}}{e^{ - \frac{x}{{\eta {\rho _e}{a_n}}}}}}}\left( {\frac{1}{{\eta {\rho _e}{a_n}{x^\delta }}} + \frac{\delta }{{{x^{\delta {\rm{ + }}1}}}}} \right){e^{ - \frac{x}{{\eta {\rho _e}{a_n}}}}}$.
\begin{remark}\label{remark3: the n-th user with ipSIC and pSIC for PD-NOMA}
Up on substituting \eqref{The asymptotic SOP of the n-th user with ipSIC for PD-NOMA} and \eqref{The asymptotic SOP of the n-th user with pSIC for PD-NOMA} into \eqref{The definition of diversity order for SOP}, the secrecy diversity orders of the $n$-th user with ipSIC/pSIC for PD-NOMA are equal to zero and one, respectively.
\end{remark}

Following the procedures similar to the approximate results in \cite[Eq. (26) and Eq. (27)]{Yue8370069Unified} and removing the ordering operation, the asymptotic SOP of the $m$-th user for CD/PD-NOMA at high SNR region are given by
\begin{align}\label{the asymptotic SOP of the m-th user pair for CD-NOMA}
&P_{m,\infty }^{CD}\left( {{R_m}} \right) = \Xi \sum\limits_{u = 1}^U {\frac{{{b_u}}}{{K!}}} \int_0^{{\tau }}  {{e^{ - \Xi \sum\limits_{i = 0}^{K - 1} {\sum\limits_{j = 0}^i {{
   i  \choose
   j
}} } \frac{\varsigma }{{i!{x^{i - j - \delta }}}}}}}  \nonumber \\
 & \times \sum\limits_{i = 0}^{K - 1} {\sum\limits_{j = 0}^i {{
   i  \choose
   j
}} \frac{\varsigma }{{i!}}\left[ {\frac{{{a_n}\left( {j - i + \delta } \right)}}{{{x^{j - i + \delta }}\left( {{a_m} - {a_n}x} \right)}}} \right.}  + \frac{{\left( {j - i + \delta } \right)}}{{{x^{j - i + \delta {\rm{ + }}1}}}} \nonumber \\
 & \left. { + \frac{{{a_m}}}{{\eta {\rho _e}{x^{j - i + \delta }}{{\left( {{a_m} - {a_n}x} \right)}^2}}}} \right]{\left( {\frac{{x{c_u}}}{{\eta \rho \left( {{a_m} - x{a_n}} \right)}}} \right)^K}dx ,
\end{align}
and
\begin{align}\label{the asymptotic SOP of the m-th user pair for PD-NOMA}
&P_{m,\infty }^{PD}\left( {{R_m}} \right) = \kappa \sum\limits_{u = 1}^U {{b_u}} \int_0^{{\tau }} {{e^{ - \frac{{\kappa {{\left( {{a_m} - {a_n}x} \right)}^\delta }}}{{{x^\delta }}}{e^{ - \frac{x}{{\eta {\rho _e}\left( {{a_m} - {a_n}x} \right)}}}}}}} \nonumber \\
&  \times {e^{ - \frac{x}{{\eta {\rho _e}\left( {{a_m} - {a_n}x} \right)}}}}\left[ {\frac{{{a_n}\delta {{\left( {{a_m} - {a_n}x} \right)}^{\delta  - 1}}}}{{{x^\delta }}} + \frac{{\delta {{\left( {{a_m} - {a_n}x} \right)}^\delta }}}{{{x^{\delta  + 1}}}}} \right. \nonumber \\
 & \left. { + \frac{{{a_m}{{\left( {{a_m} - {a_n}x} \right)}^{\delta  - 2}}}}{{\eta {\rho _e}{x^\delta }}}} \right]\frac{{x{c_n}}}{{\eta \rho \left( {{a_m} - x{a_n}} \right)}}dx,
\end{align}
respectively, where ${a_m} > x{a_n}$, $\Xi  = \delta \pi {\lambda _e}$ and $\varsigma  = {e^{\frac{{ - x}}{{\eta {\rho _e}\left( {{a_m} - {a_n}x} \right)}}}}\Gamma \left( {j + \delta } \right){\left[ {\eta {\rho _e}\left( {{a_m} - {a_n}x} \right)} \right]^{j + \delta  - i}}$.
\begin{remark}\label{remark4: the m-th user with ipSIC and pSIC for PD-NOMA}
Up on substituting \eqref{the asymptotic SOP of the m-th user pair for CD-NOMA} and \eqref{the asymptotic SOP of the m-th user pair for PD-NOMA} into \eqref{The definition of diversity order for SOP}, the secrecy diversity orders of the $m$-th user with for CD/PD-NOMA are equal to one and $K$, respectively.
\end{remark}

\begin{proposition}\label{proposition:The asymptotic SOP of the selected user pair}
The asymptotic SOP of user pairing with ipSIC/pSIC for CD/PD-NOMA are given by
\begin{align}\label{the asymptotic SOP of the selected user pair for CD-NOMA}
P_{nm,\infty }^{CD,\xi } = 1 - \left( {1 - P_{CD,n,\infty }^\xi } \right)\left( {1 - P_{m,\infty }^{CD}} \right),
\end{align}
and
\begin{align}\label{the asymptotic SOP of the selected user pair for PD-NOMA}
P_{nm,\infty }^{PD,\xi } = 1 - \left( {1 - P_{PD,n,\infty }^\xi } \right)\left( {1 - P_{m,\infty }^{PD}} \right),
\end{align}
respectively. $P_{CD,n,\infty }^{ipSIC}$, $P_{CD,n,\infty }^{pSIC}$ and $P_{m,\infty }^{CD}\left( {{R_m}} \right)$ can be obtained from \eqref{The asymptotic SOP of the n-th user with ipSIC for CD-NOMA} and \eqref{The asymptotic SOP of  the n-th user with pSIC for CD-NOMA} and \eqref{the asymptotic SOP of the m-th user pair for CD-NOMA}, respectively. $P_{PD,n,\infty }^{ipSIC}$ and $P_{PD,n,\infty }^{pSIC}$ can be obtained from \eqref{The asymptotic SOP of the n-th user with ipSIC for PD-NOMA} and \eqref{The asymptotic SOP of the n-th user with pSIC for PD-NOMA} and \eqref{the asymptotic SOP of the m-th user pair for PD-NOMA}, respectively.
\end{proposition}
\begin{remark}\label{Remark: the user pairing with ipSIC and pSIC for CD/PD-NOMA}
Upon substituting \eqref{the asymptotic SOP of the selected user pair for CD-NOMA} and \eqref{the asymptotic SOP of the selected user pair for PD-NOMA} into \eqref{The definition of diversity order for SOP}, the secrecy diversity orders of user pairing with ipSIC/pSIC for CD-NOMA and PD-NOMA are zero/$K$ and zero/one, respectively. 
\end{remark}
\subsubsection{Internal Eavesdropping Scenario}
In this case, the CDF of ${\gamma _n}$ for the $n$-th user in main channel is invariant. As a further development,
the asymptotic SOP of the $m$-th user to wiretap the $n$-th user with ipSIC/pSIC for CD-NOMA at high SNR region can be given by
\begin{align}\label{The asymptotic SOP of the m-th user to wiretap the n-th user with ipSIC for CD-NOMA}
&P_{CD,m \to n}^{ipSIC,\infty }\left( {{R_n}} \right) = \Delta \int_0^\infty  {f_{{\gamma _{{E_{m \to n}}}}}^{CD}\left( x \right)} \sum\limits_{u = 1}^U {{b_u}} \left\{ {\Omega _I^K\Gamma \left( K \right)} \right. \nonumber \\
 & - \sum\limits_{i = 0}^{K - 1} {\frac{{{{\left( {\varpi \rho {\vartheta _1}} \right)}^i}}}{{i!}}} \left. {{{\left( {\frac{{{\Omega _I}}}{{\varpi \rho {\vartheta _1}{\Omega _I} + 1}}} \right)}^{K + i}}\Gamma \left( {K + i} \right)} \right\}dx,
\end{align}
and
\begin{align}\label{The asymptotic SOP of the m-th user to wiretap the n-th user with pSIC for CD-NOMA}
P_{CD,m \to n}^{pSIC,\infty }\left( {{R_n}} \right) = \Delta \sum\limits_{u = 1}^U {\frac{{{b_u}}}{{K!}}} \int_0^\infty  {\vartheta _1^Kf_{{\gamma _{{E_{m \to n}}}}}^{CD}\left( x \right)} dx,
\end{align}
where ${f_{{\gamma _{{E_{m \to n}}}}}^{CD}}$ can be obtained from \eqref{the PDF of far user for CD-NOMA in IE case}.
\begin{remark}\label{remark4: the n-th user with ipSIC and pSIC for PD-NOMA}
Up on substituting \eqref{The asymptotic SOP of the m-th user to wiretap the n-th user with ipSIC for CD-NOMA} and \eqref{The asymptotic SOP of the m-th user to wiretap the n-th user with pSIC for CD-NOMA} into \eqref{The definition of diversity order for SOP}, the secrecy diversity orders of the $m$-th user to wiretap the $n$-th user with ipSIC/pSIC for CD-NOMA are equal to zero and $K$, respectively.
\end{remark}

The asymptotic SOP of the $m$-th to wiretap the $n$-th user with ipSIC/pSIC for PD-NOMA at high SNR region can be given by
\begin{align}\label{The asymptotic SOP of the m-th user to wiretap the n-th user with ipSIC for PD-NOMA}
P_{PD,m \to n}^{ipSIC,\infty }\left( {{R_n}} \right) = \sum\limits_{u = 1}^U {{b_u}} \int_0^\infty  {\mu {e^{ - \frac{\mu }{{{x^\delta }}}{e^{ - \frac{x}{{\eta \rho_{e} {a_n}}}}} - \frac{x}{{\eta \rho_{e} {a_n}}}}}}  \nonumber \\
  \times \left( {\frac{1}{{\eta \rho_{e} {a_n}{x^\delta }}} + \frac{\delta }{{{x^{\delta {\rm{ + }}1}}}}} \right)\left( {\frac{{{\vartheta _1} \rho {\Omega _I}}}{{1 + {\vartheta _1} \rho {\Omega _I}}}} \right)dx ,
\end{align}
and
\begin{align}\label{The asymptotic SOP of the m-th user to wiretap the n-th user with pSIC for PD-NOMA}
P_{PD,m \to n}^{pSIC,\infty }\left( {{R_n}} \right) = & \sum\limits_{u = 1}^U {{b_u}} \int_0^\infty  {\mu {\vartheta _1} {e^{ - \frac{\mu }{{{x^\delta }}}{e^{ - \frac{x}{{\eta \rho_{e} {a_n}}}}} - \frac{x}{{\eta \rho_{e} {a_n}}}}}}  \nonumber \\
&  \times \left( {\frac{1}{{\eta \rho_{e} {a_n}{x^\delta }}} + \frac{\delta }{{{x^{\delta {\rm{ + }}1}}}}} \right)dx ,
\end{align}
respectively.

\begin{remark}\label{remark5: the n-th user with ipSIC and pSIC for PD-NOMA}
Up on substituting \eqref{The asymptotic SOP of the m-th user to wiretap the n-th user with ipSIC for PD-NOMA} and \eqref{The asymptotic SOP of the m-th user to wiretap the n-th user with pSIC for PD-NOMA} into \eqref{The definition of diversity order for SOP}, the secrecy diversity orders of the $m$-th user to wiretap the $n$-th user with ipSIC/pSIC for PD-NOMA are equal to zero and one, respectively.
\end{remark}

From the above remarks, we observe that the secrecy diversity orders of LUs are not only related to the number of subcarriers $K$, but also affected by the residual interference from SIC. Hence it is important to consider the size of subcarriers $K$ and residual interference in realistic secure communication scenarios.

\begin{table}[!t]
\centering
\caption{Table of Parameters for Numerical Results}
\tabcolsep5pt
\renewcommand\arraystretch{1.1} 
\begin{tabular}{|l|l|}
\hline
Monte Carlo simulations repeated  &  ${10^5}$ iterations \\
\hline
Carrier frequency  &  $1 $ GHz  \\
\hline
The radius of a disc region for Eves  &  1000 m \\
\hline
Path loss exponent   & $\alpha=2$  \\
\hline
\multirow{1}{*}{Power allocation coefficients of NOMA} &  \multirow{1}{*}{ $a_m=0.8$, $a_n=0.2$}   \\
\hline
\multirow{2}{*}{Targeted data rates}  & \multirow{1}{*}{$R_{{n}}=R_{{m}}=0.01 $ BPCU}  \\
                                      & \multirow{1}{*}{$R_{{m \to n}}=0.01$ BPCU}  \\
\cline{1-2}
\multirow{1}{*}{The radius of the $n$-th LU zone}  &  $R_{D_1}=2$ m \\
\hline
The radius of the $m$-th LU zone  &  $R_{D_2}=10$ m \\
\hline
\end{tabular}
\label{parameter}
\end{table}
\section{Numerical Results}\label{Numerical Results}
In this section, numerical results of the secrecy transmission for the unified NOMA framework are present to verify the accuracy of the analytical expressions, where both external and internal eavesdropping scenarios are discussed in detail. We show interplay of different system configuration parameters and impacts on the SOP for CD/PD-NOMA networks. Monte Carlo simulation parameters used in this section are summarized in Table~\ref{parameter}, where BPCU is abbreviation of bit per channel use. Simulation results are denoted by $ \bullet $ and the complexity-vs-accuracy tradeoff parameter is set to be $U=15$. In external eavesdropping scenario, the conventional OMA scheme is selected to be a benchmark for comparing the secrecy performance of unified NOMA framework, while the security performance of PD-NOMA is viewed as baseline for the purpose of comparison in internal eavesdropping scenario.
\subsection{External Eavesdropping Scenario}
In this subsection, the secrecy performance of CD/PD-NOMA networks is characterized in terms of SOP for external  internal eavesdropping scenario.
\begin{figure}[t!]
    \begin{center}
        \includegraphics[width=3.3in,  height=2.4in]{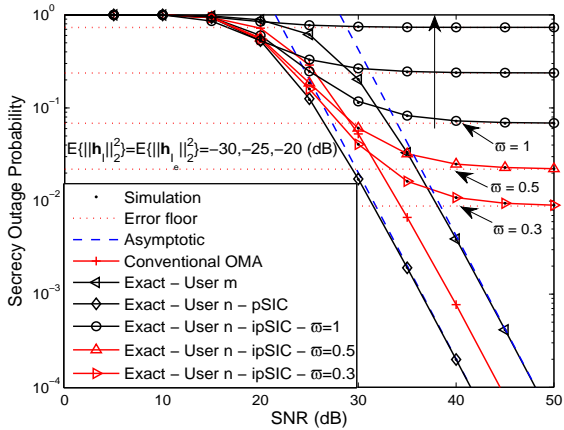}
        \caption{The SOP versus the transmit SNR, with ${\rho _e} = 10$ dB, ${\lambda _e} = {10^{ - 3}}$, $K=2$, $\alpha = 2$ and $R_n$ = $R_m$ = 0.01 BPCU.}
        \label{The_SOP_EE_diff_IR}
    \end{center}
\end{figure}

Fig. \ref{The_SOP_EE_diff_IR} plots the SOP versus the transmit SNR with different residual interference values, where $K = 2$, $\alpha = 2$ and $R_n$ = $R_m$ = 0.01 BPCU. The exact analytical curves of SOP for the $n$-th user with ipSIC and pSIC are plotted based on \eqref{the expression for the n-th user with ipSIC for CD-NOMA} and \eqref{CD-NOMA:the expression for the m-th user with pSIC}, respectively. The exact theoretical curve of SOP for the $m$-th user is plotted based on \eqref{the SOP of the m-th user for CD-NOMA}. It can be seen that the numerical evaluation curves match with analytical results. The exact asymptotic curves of SOP for the $n$-th user with ipSIC and pSIC are plotted based on \eqref{The asymptotic SOP of the n-th user with ipSIC for CD-NOMA} and \eqref{The asymptotic SOP of the n-th user with pSIC for CD-NOMA}, respectively. The exact asymptotic curve of SOP for the $m$-th user is plotted based on \eqref{the asymptotic SOP of the m-th user pair for CD-NOMA}. Moreover, the approximate analysis results of SOP perfectly match with the theoretical analysis in the high SNR region. One can observed that the secrecy outage behavior of the $n$-th user with pSIC is superior to that of the OMA scheme, while the security performance of the $m$-th user is inferior to OMA. This comes from the fact that NOMA scheme is capable of providing more user fairness in comparison to OMA \cite{Ding2014performance,Men7454773TVT,Yue7812773Fixed}.
Another significant observation is that SOP of the $n$-th user with ipSIC converge to an error floor and thus gain a zero secrecy diversity order. The reason is that the ipSIC scheme employed at $n$-th user for the main channel brings the effect of residual interference on outage probability, which verifies the conclusion in \textbf{Remark \ref{remark1: the n-th user with ipSIC for CD-NOMA}}.
In the case of fixed the residual interference level, i.e., $\varpi = 1$, as the residual interference value from $-30$ dB to $-20$ dB increases, the outage performance of the $n$-th user with ipSIC is becoming more worse. That is to say that the corresponding error floors for SOP are becoming more larger.
Furthermore, the different impact levels of residual interference also degrade the secrecy performance seriously. More specifically, under the condition of ${\rm{E}}\left\{ \| {\mathbf{h_I}} \right\|_2^2 \} = {\rm{E}}\left\{ \| {\mathbf{h_{{I_e}}}\|_2^2} \right\} =  - 30$ dB, with the value of $\varpi$ increasing from ${\varpi} = 0.3$ to ${\varpi} = 1$, the superiority of outage behaviors for the $n$-th user with ipSIC is no longer obvious. These imply how to eliminate the residual interference should be taken into account in practical secrecy NOMA scenarios.

\begin{figure}[t!]
    \begin{center}
        \includegraphics[width=3.3in,  height=2.4in]{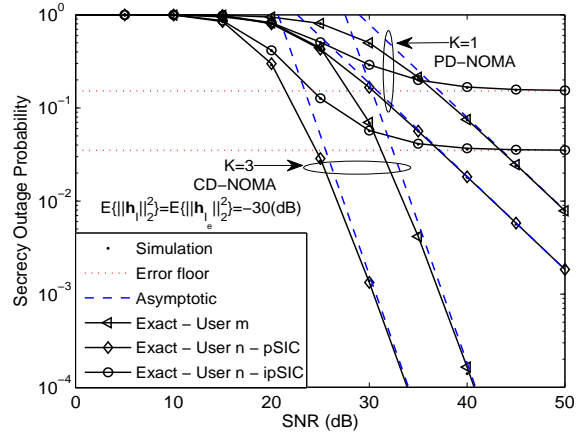}
        \caption{The SOP versus the transmit SNR, with ${\rho _e} = 10$ dB, ${\lambda _e} = {10^{ - 3}}$, $\alpha = 2$, $\varpi {\rm{ =  1}}$ and $R_n$ = $R_m$ = 0.01 BPCU.}
        \label{The_SOP_EE_diff_K}
    \end{center}
\end{figure}
Fig. \ref{The_SOP_EE_diff_K} plots the SOP versus the transmit SNR with different subcarriers $K$, ${\rho _e} = 10$ dB, ${\lambda _e} = {10^{ - 3}}$, $\alpha = 2$,  $\varpi {\rm{ =  1}}$ and $R_n$ = $R_m$ = 0.01 BPCU. As can be seen from this figure, with the number of subcarriers increasing, the slopes of SOP for CD-NOMA $(K=3)$ are larger than that of PD-NOMA $(K=1)$.  This can be explained by the fact that the secrecy CD-NOMA network provides a diversity gain that is equal to the number of subcarriers $K$, which is also validated by the insights in \textbf{Remark \ref{remark3: the n-th user with ipSIC and pSIC for PD-NOMA}} and \textbf{Remark \ref{remark4: the m-th user with ipSIC and pSIC for PD-NOMA}}. This implies that by spreading the user's information into a plurality of subcarriers, CD-NOMA has enhanced the secrecy performance of users achieved and provided spread spectrum gain with respect to PD-NOMA.

\begin{figure}[t!]
    \begin{center}
        \includegraphics[width=3.3in,  height=2.4in]{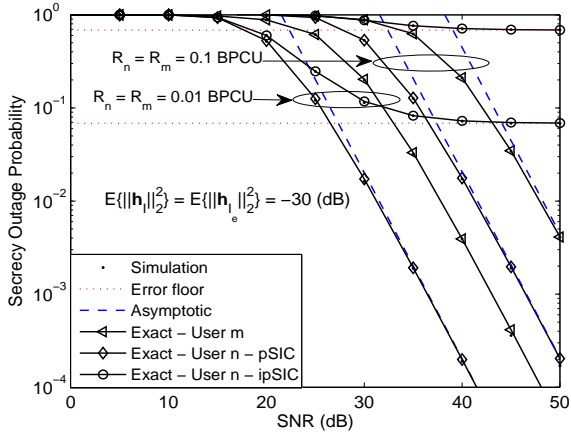}
        \caption{The SOP versus the transmit SNR, with ${\rho _e} = 10$ dB, ${\lambda _e} = {10^{ - 3}}$, $K=2$, $\alpha = 2$ and $\varpi {\rm{ =  1}}$.}
        \label{The_SOP_diff_Target_rate}
    \end{center}
\end{figure}
Fig. \ref{The_SOP_diff_Target_rate} plots the SOP versus the transmit SNR with different target secrecy rates, where ${\rho _e} = 10$ dB, ${\lambda _e} = {10^{ - 3}}$, $K=2$, $\alpha = 2$ and  $\varpi {\rm{ =  1}}$. One can show that the selection of target secrecy rate has a greater impact on the SOP for NOMA networks. With increasing of the target secrecy rates, the outage probability of users for secrecy NOMA are becoming worse seriously. It is worth pointing out that the adaptive secrecy target rates has a large effect on the performance of secure communication for CD/PD-NOMA networks.
As a consequence, the requirements of smaller secrecy rate can be applied into small packet business and internet of things scenarios.

\begin{figure}[t!]
    \begin{center}
        \includegraphics[width=3.3in,  height=2.4in]{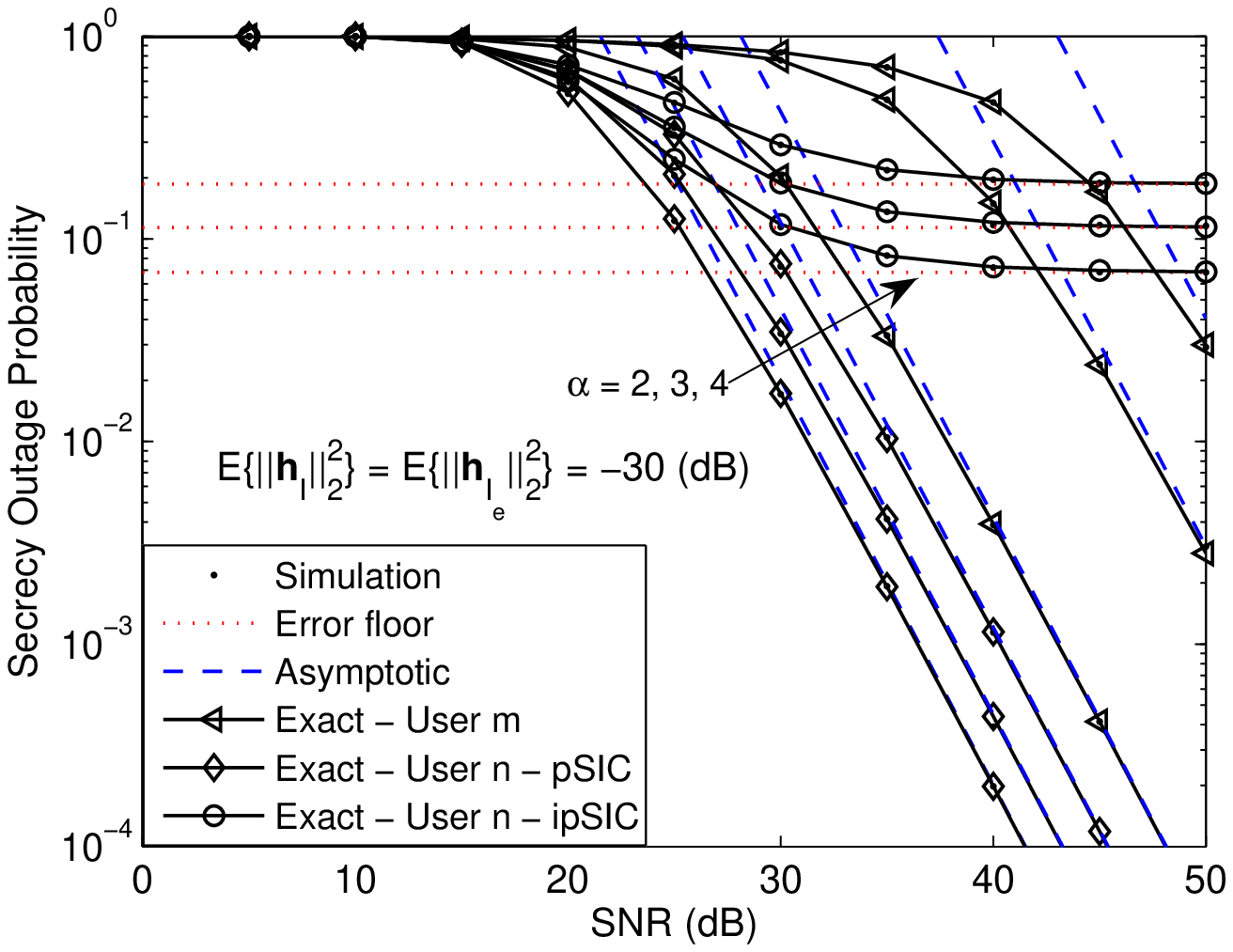}
        \caption{The SOP versus the transmit SNR, with ${\rho _e} = 10$ dB, ${\lambda _e} = {10^{ - 3}}$, $K=2$ and $\varpi {\rm{ =  1}}$.}
        \label{The_SOP_diff_alpha}
    \end{center}
\end{figure}
\begin{figure}[t!]
    \begin{center}
        \includegraphics[width=3.3in,  height=2.4in]{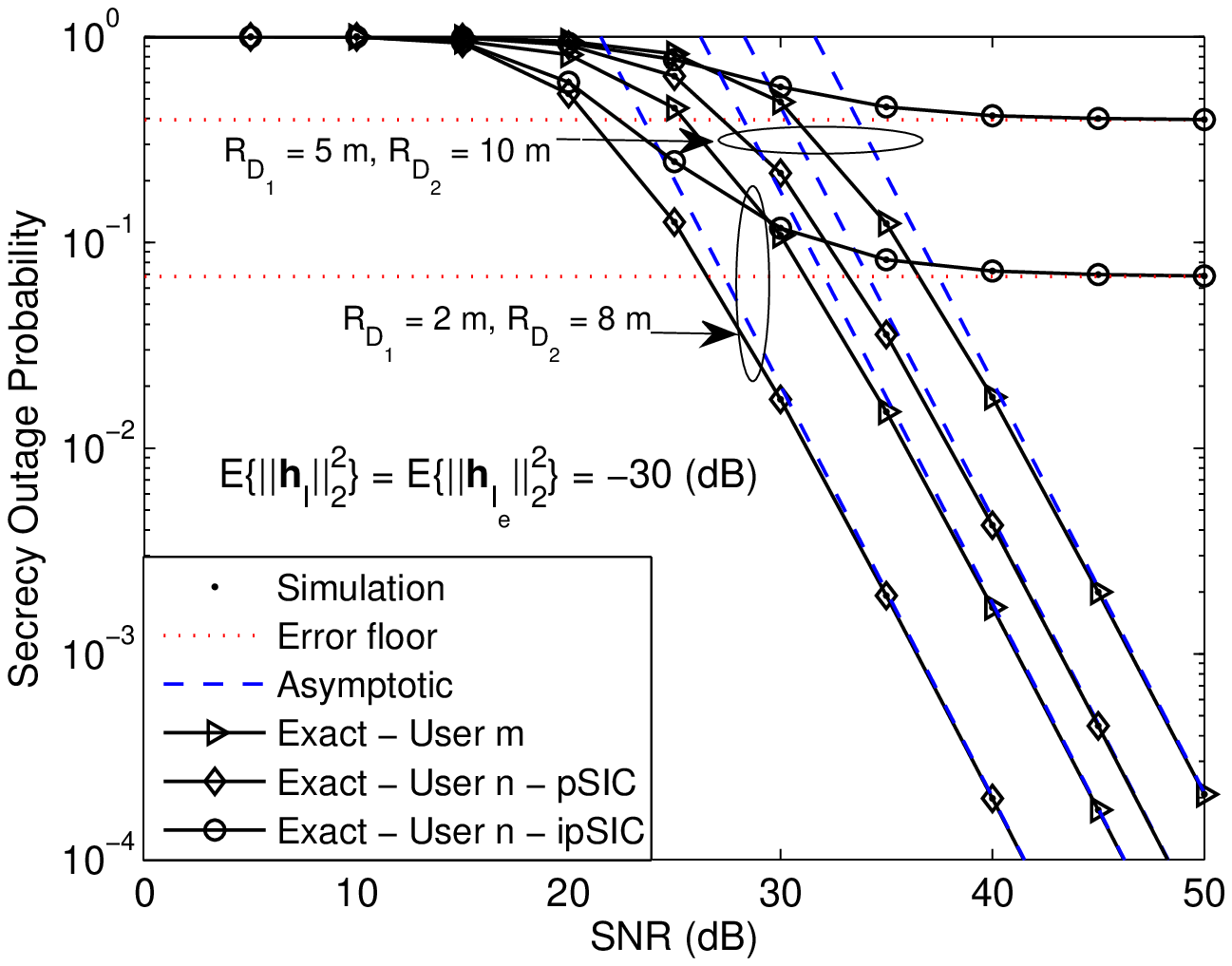}
        \caption{The SOP versus the transmit SNR, with ${\rho _e} = 10$ dB, ${\lambda _e} = {10^{ - 3}}$, $K=2$ and  $\varpi {\rm{ =  1}}$.}
        \label{The_SOP_EE_n_m_diff_distance}
    \end{center}
\end{figure}

Fig. \ref{The_SOP_diff_alpha} plots the SOP versus the transmit SNR with different path loss exponents $\alpha = 2, 3, 4$, where ${\rho _e} = 10$ dB, ${\lambda _e} = {10^{ - 3}}$, $K=2$ and  $\varpi {\rm{ =  1}}$. It can be seen from this figure that by increasing the path loss exponent, i.e., more severe path loss, leads to a poorer secrecy outage behaviors. It is worth mentioning that the effect of path loss on the secrecy performance of the $n$-th user is smaller, while has a greater impact on the $m$-th user. The main reason for this result is that the $m$-th user is far away from BS and is greatly affected by large-scale fading. Furthermore, the SOP versus the transmit SNR with different radius distance is plotted in Fig. \ref{The_SOP_EE_n_m_diff_distance}. One can observe that the better SOP can be achieved by reducing the radius of the user distance, since the smaller radius distance results in a lower path loss.

\begin{figure}[t!]
    \begin{center}
        \includegraphics[width=3.3in,  height=2.4in]{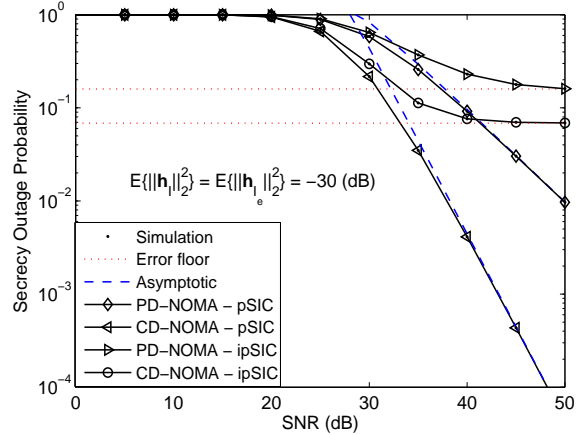}
        \caption{The SOP of user pairing versus the transmit SNR, with ${\rho _e} = 10$ dB, ${\lambda _e} = {10^{ - 3}}$, $K=2$ and  $\varpi {\rm{ =  1}}$.}
        \label{The_SOP_User_Pairing}
    \end{center}
\end{figure}
Fig. \ref{The_SOP_User_Pairing} plots the SOP of a pair of NOMA users (the $n$-th user and the $m$-th user) versus the transmit SNR, where ${\rho _e} = 10$ dB, ${\lambda _e} = {10^{ - 3}}$, $K=2$ and  $\varpi {\rm{ =  1}}$. The exact analytical curves of SOP for CD/PD-NOMA networks are plotted based on \eqref{the SOP of the selected user pair for CD-NOMA} and \eqref{the SOP of the selected user pair for PD-NOMA}, respectively. It can be seen that simulation results well match the theoretical results perfectly.
The asymptotic curves of SOP for CD/PD-NOMA networks are plotted based on \eqref{the asymptotic SOP of the selected user pair for CD-NOMA} and \eqref{the asymptotic SOP of the selected user pair for PD-NOMA}, respectively.
The approximate results converge to exact analytical results at high SNR. One can observe that the secrecy performance of CD-NOMA is superior to that of PD-NOMA, since CD-NOMA is capable of providing a larger secrecy diversity order. This phenomenon also verify the conclusion in \textbf{Remark \ref{Remark: the user pairing with ipSIC and pSIC for CD/PD-NOMA}}.
Another observation is that in the absence of channel ordering between BS and users, the secrecy performance of CD/PD-NOMA is not depend on that of the distant user.

\begin{figure}[t!]
    \begin{center}
        \includegraphics[width=3.3in,  height=2.4in]{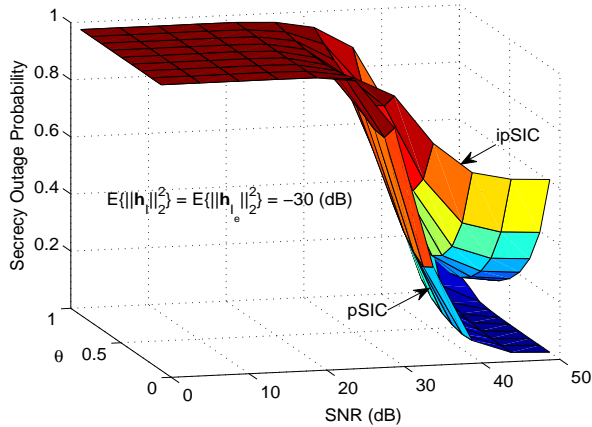}
        \caption{The SOP of user pairing versus the transmit SNR, with ${\rho _e} = 10$ dB, ${\lambda _e} = {10^{ - 3}}$, $K=2$ and $\varpi {\rm{ =  1}}$.}
        \label{The_SOP_User_Pairing_diff_3D}
    \end{center}
\end{figure}
To observe how the power allocation factor affects the secrecy outage performance in the unified NOMA framework, we present the curves of SOP versus dynamic power factor $\theta$ in Fig. \ref{The_SOP_User_Pairing_diff_3D}, where $a_n = \theta$ and $a_m = 1-  \theta$. As can observed from the figure, with the increase of SNR, the SOP of user pairing is gradually reduced on the condition of pSIC, while the SOP of user pairing with ipSIC converges to a constant value. This phenomenon can be explained as that the imperfect cancellation scheme employed brings more residual interference. Fixing the SNR, i.e., $\rho = 40$ dB, when $\theta$ decreases, the SOP of user pairing with pSIC is becoming smaller; On the contrary, the change degrades the secrecy performance with ipSIC. This is due to the fact that the signal power of the nearby user mainly affect the secure performance.
As a consequence, the adaptive power allocation factors between NOMA users are important for achieving the optimal security performance.
\subsection{Internal Eavesdropping Scenario}
In this subsection, the secrecy outage behaviors of the $m$-th user to wiretap the $n$-th user are evaluated for CD/PD-NOMA.

\begin{figure}[t!]
    \begin{center}
        \includegraphics[width=3.3in,  height=2.4in]{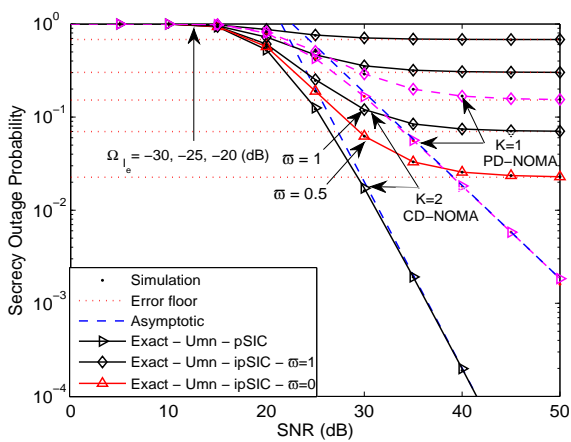}
        \caption{The SOP versus the transmit SNR, with ${\rho _e} = 10$ dB, ${\lambda _e} = {10^{ - 3}}$, $K=2$ and $R_{D_1}=2$ m.}
        \label{The_SOP_IE_K2_diff_IR_diff_K}
    \end{center}
\end{figure}
Fig. \ref{The_SOP_IE_K2_diff_IR_diff_K} plots the SOP versus the transmit SNR, where ${\rho _e} = 10$ dB, ${\lambda _e} = {10^{ - 3}}$, $K=2$ and $R_{D_1}=2$ m.
Under the condition of ipSIC/pSIC, the exact SOP curves of the $m$-th user to wiretap the $n$-th user for CD-NOMA and PD-NOMA are
plotted based on \eqref{the SOP of the m-th user to wiretap the n-th user with ipSIC for CD-NOMA}, \eqref{the SOP of the m-th user to wiretap the n-th user with pSIC for CD-NOMA} and \eqref{the SOP of the m-th user to wiretap the n-th user with ipSIC for PD-NOMA},  \eqref{the SOP of the m-th user to wiretap the n-th user with pSIC for PD-NOMA}, respectively. The analytical results are consistent with simulation results, which further confirm the accuracy of derived results. As can be observed from this figure that the secrecy outage behaviors of the $m$-th user to wiretap the $n$-th user for CD-NOMA outperform that of the $m$-th user to wiretap the $n$-th user for PD-NOMA. This is because that CD-NOMA is capable of obtaining more diversity gain.
Another observation is that the residual interference levels of ipSIC have a greater impact on secrecy behaviors. As $\varpi$ becomes small from $\varpi {\rm{ =  1}}$ to $\varpi {\rm{ =  0.5}}$, the performance gain brought by ipSIC is becoming more obvious.
Additionally, increasing the value of residual interference with the fixed interference level, i.e., $\varpi {\rm{ =  1}}$, the secrecy behaviors of internal eavesdropping scenario are becoming worse. Hence it is important to premeditate the influence of residual interference  when performing physical layer secure communication.

\begin{figure}[t!]
    \begin{center}
        \includegraphics[width=3.3in,  height=2.4in]{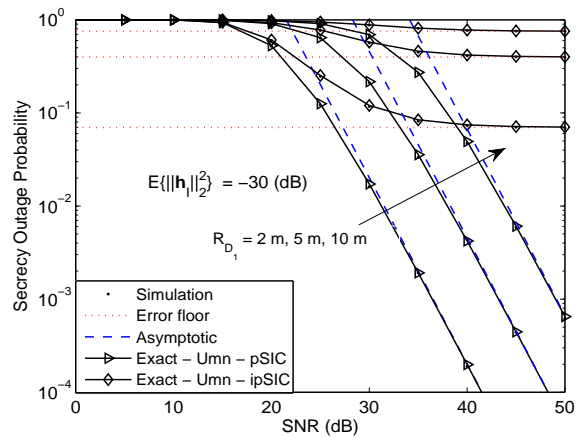}
        \caption{The SOP versus the transmit SNR, with ${\rho _e} = 10$ dB, ${\lambda _e} = {10^{ - 3}}$, $K=2$ and $\varpi {\rm{ =  1}}$.}
        \label{The_SOP_IE_diff_distance}
    \end{center}
\end{figure}
Fig. \ref{The_SOP_IE_diff_distance} plots the SOP versus the transmit SNR with different distances, where ${\rho _e} = 10$ dB, ${\lambda _e} = {10^{ - 3}}$, $K=2$,  $\varpi {\rm{ =  1}}$ and $\mathbb{E} \{ {\left\| {{{\bf{h}}_I}} \right\|_2^2} \}= -30$ dB. It can be obviously seen from the figure that the theoretical and simulation results are closely matched for different parameters. As can be observed from figure that increasing the distance between BS and users will reduce the SOP, which can be explained as the secrecy behaviors of the $m$-th user to wiretap the $n$-th user are progressively affected by the large scale fading.
\section{Conclusion}\label{Conclusion}
In this paper, the physical layer security issues are discussed exhaustively in a unified NOMA framework, where both external and internal eavesdropping scenarios are taken into account. The spatial locations of LUs and Eves are modeled by applying statistic geometry. The secrecy behavior of a unified NOMA framework has been investigated insightfully.
New exact and asymptotic expressions of SOP for CD/PD-NOMA with ipSIC/pSIC have been derived to characterize the secrecy performance. One can observe that the secrecy diversity orders of the $m$-th user for CD/PD-NOMA are $K$ and one, respectively. Due to the effect of residual interference, a zero secrecy diversity order has been obtained by the $n$-th user with ipSIC for both eavesdropping scenarios in CD/PD-NOMA networks. Based on analytical results, it was shown that the secrecy diversity orders of LUs with pSIC for CD-NOMA are depended on the number of subcarries considered in the networks.
Finally, numerical simulations were presented to verify the accuracy of analytical results.

\appendices
\section*{Appendix~A: Proof of Lemma \ref{Lemmad: The_CDF of near LUs with ipSIC for CD-NOMA}} \label{Appendix:A}
\renewcommand{\theequation}{A.\arabic{equation}}
\setcounter{equation}{0}

The proof starts by assuming ${{\bf{G}}_{K \times M}}$ belongs to regular sparse matrix with ${{{\bf{g}}_n}}$ and ${{{\bf{g}}_m}}$ having the same column weights. Based on \eqref{the SINR expression at n-th LU to detect itself with SIC}, the CDF $F_{CD,{\gamma _n}}^{ipSIC}$ of ${{\gamma _n}}$ for CD-NOMA can be expressed as
\begin{align}\label{Appenxix A: The CDF of n-th user with ipSIC for CD-NOMA}
F_{CD,{\gamma _n}}^{ipSIC}\left( x \right) = {\rm{Pr}}\left( {\frac{{\rho \left\| {diag\left( {{{\bf{h}}_n}} \right){{\bf{g}}_n}} \right\|_2^2{a_n}}}{{\varpi \rho \left\| {{{\bf{h}}_I}} \right\|_2^2 + 1}} < x} \right),
\end{align}
where $\varpi \ne 0$.
By the virtue of \cite[Eq. (8)]{Yue8370069Unified} and removing ordering operation, the equation \eqref{Appenxix A: The CDF of n-th user with ipSIC for CD-NOMA} can be further calculated as follows:
\begin{align}\label{Appenxix AA: The CDF of n-th user with ipSIC for CD-NOMA}
&F_{CD,{\gamma _n}}^{ipSIC}\left( x \right) \approx \Delta \int_0^\infty  {{y^{K - 1}}{e^{ - \frac{y}{{{\Omega _I}}}}}} \left\{ {\sum\limits_{u = 1}^U {{b_u}} } \right.\left[ 1 \right. \nonumber\\
 &\left. {\left. { - {e^{ - \frac{{x{c_u}\left( {\varpi \rho y + 1} \right)}}{{\eta \rho {a_n}}}}}\sum\limits_{i = 0}^{K - 1} {\frac{1}{{i!}}{{\left( {\frac{{x{c_u}\left( {\varpi \rho y + 1} \right)}}{{\eta \rho {a_n}}}} \right)}^i}} } \right]} \right\}dy \\
  & = \Delta \sum\limits_{u = 1}^U {{b_u}} \left[ {\underbrace {\int_0^\infty  {{y^{K - 1}}{e^{ - \frac{y}{{{\Omega _I}}}}}} dy}_{{J_1}}} \right. - {e^{ - \frac{{x{c_u}}}{{\eta \rho {a_n}}}}}\sum\limits_{i = 0}^{K - 1} {\frac{1}{{i!}}{{\left( {\frac{{x{c_u}}}{{\eta \rho {a_n}}}} \right)}^i}} \nonumber \\
  & { \times \underbrace {\int_0^\infty  {{y^{K - 1}}{e^{ - \frac{{y\left( {x\varpi {c_u}{\Omega _I}{\rm{ + }}\eta {a_n}} \right)}}{{\eta {a_n}{\Omega _I}}}}}} {{\left( {\varpi \rho y + 1} \right)}^i}dy}_{{J_2}}} ,
\end{align}
where $\Delta  = \frac{1}{{\left( {K - 1} \right){\rm{!}}\Omega _I^K}}$.

By the virtue of \cite[Eq. (3.381.4)]{gradshteyn} and using Binomial theorem, $J_1$ and $J_2$ are given by
\begin{align}\label{J1}
J_1 = \Omega _I^K\Gamma \left( K \right),
\end{align}
and
\begin{align}\label{J2}
{J_2} =& {e^{ - \frac{{x{c_u}}}{{\eta \rho {a_n}}}}}\sum\limits_{i = 0}^{K - 1} {\sum\limits_{j = 0}^i {\frac{1}{{i!}}{
   i  \choose
   j
}{{\left( {\frac{{x{c_u}}}{{\eta \rho {a_n}}}} \right)}^i}} }  \nonumber \\
  & \times {\left( {\frac{{\eta {a_n}{\Omega _I}}}{{x\varpi {c_u}{\Omega _I} + \eta {a_n}}}} \right)^{K + j}}{\left( {\varpi \rho } \right)^j}\Gamma \left( {K + j} \right),
\end{align}
respectively.

Upon substituting \eqref{J1} and \eqref{J2} into (A.3), we can obtain \eqref{CD-NOMA:the CDF of SINR expression for the near LUs with ipSIC} and complete the proof.

\appendices
\section*{Appendix~B: Proof of Lemma \ref{CD-NOMA:the_PDF for near user with ipSIC}} \label{Appendix:B}
\renewcommand{\theequation}{B.\arabic{equation}}
\setcounter{equation}{0}
To derive the PDF ${f_{CD,{\gamma _{{E_n}}}}^{ipSIC}}$ of the most pernicious Eve, we should first calculate the corresponding CDF ${F_{CD,{\gamma _{{E_n}}}}^{ipSIC}}$ previously. For notational simplicity, define $X \buildrel \Delta \over = \left\| {diag\left( {{{\bf{h}}_e}} \right){{\bf{g}}_e}} \right\|_2^2 = {\frac{{\eta Y}}{{1 + {d_e^\alpha }}}}$, $Y = \sum\limits_{k = 1}^K {{{\left| {{g_{ek}}{{\tilde h}_{ek}}} \right|}^2}}$ and $Z\buildrel \Delta \over = \left\| {{{\bf{h}}_{{I_e}}}} \right\|_2^2$.
It may be readily observe that $Y$ and $Z$ obeys a Gamma distribution with the parameters of $\left( {K,1} \right)$ and $\left( {K,{\Omega _{{I_e}}}} \right)$, respectively.
Hence the CDF and PDF of $Y$ and $Z$ are given by
${F_Y}\left( y \right){\rm{ = }}1 - {e^{ - y}}\sum\limits_{i = 0}^{K - 1} {\frac{{{y^i}}}{{i!}}}$ and ${f_Z}\left( z \right) = \frac{{{y^{K - 1}}{e^{ - \frac{z}{{{\Omega _{{I_e}}}}}}}}}{{\Omega _{{I_e}}^K\left( {K - 1} \right)}}$, respectively.

Based on \eqref{the SINR expression of eavesdropper to detect the n-th user}, the CDF ${F_{CD,{\gamma _{{E_n}}}}^{ipSIC}}$ of ${\gamma _{{E_n}}}$ for CD-NOMA can be expressed as
\begin{align}\label{Appenxix: The CDF of Eves to wiretap the n-th user for CD-NOMA}
&F_{CD,{\gamma _{{E_n}}}}^{ipSIC} (x)= {\rm{Pr}}\left[ {\mathop {\max }\limits_{e \in {\Phi _e}} \left( {\frac{{{\rho _e}\left\| {diag\left( {{{\bf{h}}_e}} \right){{\bf{g}}_e}} \right\|_2^2{a_n}}}{{\varpi {\rho _e}\left\| {{{\bf{h}}_{{I_e}}}} \right\|_2^2 + 1}}} \right) < x} \right]  \nonumber \\
&= {{\mathop{\rm \mathbb{E}}\nolimits} _{{\Phi _e}}}\left\{ {\prod\limits_{e \in {\Phi _e}} {\int_0^\infty  {{f_Z}\left( z \right){F_Y}\left[ {\frac{{x\left( {\varpi {\rho _e}z + 1} \right)\left( {1 + d_e^\alpha } \right)}}{{{\varphi _1}}}} \right]} dz} } \right\},
\end{align}
where $\varpi \ne 0$.

By applying the generating function \cite{chiu2013stochastic} and polar coordinate conversion, the equation \eqref{Appenxix: The CDF of Eves to wiretap the n-th user for CD-NOMA} can be calculated as follows:
\begin{align}\label{the CDF of EE with ipSIC for CD-NOMA}
&F_{CD,{\gamma _{{E_n}}}}^{ipSIC}\left( x \right) =  \exp \left\{ { - 2\pi {\lambda _e}\int_0^\infty  {\left[ {1 - \int_0^\infty  {{f_Z}\left( z \right)} } \right.} } \right. \nonumber\\
 &\left. { \times \left. {{F_Y}\left( {\frac{{x\left( {\varpi {\rho _e}z + 1} \right)\left( {1 + {r^\alpha }} \right)}}{{{\varphi _1}}}} \right)dz} \right]rdr} \right\}  \nonumber\\
  & = \exp \left\{ { \Phi \int_0^\infty  {\frac{{{e^{ - \frac{{x\left( {1 + {r^\alpha }} \right)}}{{{\varphi _1}}}}}}}{{\left( {K - 1} \right)!\Omega _{{I_e}}^K}}\sum\limits_{i = 0}^{K - 1} {\sum\limits_{j = 0}^i {{
   i  \choose
   j
}} \frac{{{{\left( {\varpi {\rho _e}} \right)}^j}}}{{i!}}} } } \right. {{\left( {\frac{x}{{{\varphi _1}}}} \right)}^i} \nonumber \\
  &  \left. { \times {{\left( {1 + {r^\alpha }} \right)}^i}r\underbrace {\int_0^\infty  {{z^{K + j - 1}}{e^{ - \frac{{z\left[ {{\varphi _1} + x\varpi \left( {1 + {r^\alpha }} \right){\rho _e}{\Omega _{{I_e}}}} \right]}}{{{\varphi _1}{\Omega _{{I_e}}}}}}}} dz}_{{J_3}}dr} \right\},
\end{align}
where ${\varphi _1}{\rm{ = }}\eta {\rho _e}{a_n}$ and $\Phi {\rm{ = }} - 2\pi {\lambda _e}$.

By the virtue of  \cite[Eq. (3.351.3)]{gradshteyn}, $J_3$ is given by
\begin{align}\label{J3}
{J_3} = \left( {K + j - 1} \right)!{\left( {\frac{{{\varphi _1} + x\varpi \left( {1 + {r^\alpha }} \right){\rho _e}{\Omega _{{I_e}}}}}{{{\varphi _1}{\Omega _{{I_e}}}}}} \right)^{ - K - j}},
\end{align}

Upon substituting \eqref{J3} into \eqref{the CDF of EE with ipSIC for CD-NOMA}, the CDF ${F_{CD,{\gamma _{{E_n}}}}^{ipSIC}}$ is given by
\begin{align}\label{Appenxix: The CDF of Eves to wiretap the n-th user for CD-NOMA final expression}
&F_{CD,{\gamma _{{E_n}}}}^{ipSIC}\left( x \right) = \exp \left\{ {\Phi \sum\limits_{i = 0}^{K - 1} {\sum\limits_{j = 0}^i {\frac{{{x^i}\phi \zeta }}{{i!}}} } } \right.\int_0^\infty  {{{\left( {1 + {r^\alpha }} \right)}^i}}  \nonumber\\
 & \times \left. {{{\left[ {{\varphi _1} + x\varpi {\varphi _2}\left( {1 + {r^\alpha }} \right)} \right]}^{ - K - j}}{e^{ - \frac{{x\left( {1 + {r^\alpha }} \right)}}{{{\varphi _1}}}}}rdr} \right\},
\end{align}
where ${\varphi _2}{\rm{ = }}{\rho _e}{\Omega _{{I_e}}}$, $\phi = \frac{{i!}}{{\left( {i - j} \right)!j!}}$ and $\zeta  = \frac{{{{\left( {\varpi {\varphi _2}} \right)}^j}\left( {K + j - 1} \right)!}}{{\varphi _1^{ - K - j + i}\left( {K - 1} \right)!}}$.

After that, applying the derivative of $F_{CD,{\gamma _{{E_n}}}}^{ipSIC}$ and some manipulations, the PDF ${f_{CD,{\gamma _{{E_n}}}}^{ipSIC}}$ can be obtain in \eqref{CD-NOMA:the PDF expression for the Eves with ipSIC} and the proof is completed.

\appendices
\section*{Appendix~C: Proof of Corollary  \ref{Corollary:PD-NOMA:the COP of far user}} \label{Appendix:C}
\renewcommand{\theequation}{C.\arabic{equation}}
\setcounter{equation}{0}
Denoting $\tilde X \buildrel \Delta \over = {\left| {{{\tilde h}_{ek}}} \right|^2} = \frac{{\eta \tilde Y}}{{1 + d_e^\alpha }} $, $\tilde Y = {\left| {{h_{ek}}} \right|^2}$ and $\tilde Z = {\left| {{h_{{I_e}k}}} \right|^2}$, we can observe that the CDF and PDF of $\tilde Y$ and $\tilde Z$ are equal to ${F_{\tilde Y}}\left( y \right){\rm{ = }}1 - {e^{ - y}}$ and ${f_{\tilde Z}}\left( z \right) = \frac{1}{{{\Omega _{{I_e}}}}}{e^{ - \frac{z}{{{\Omega _{{I_e}}}}}}}$, respectively.
Based on \eqref{the SINR expression of eavesdropper to detect the n-th user}, the CDF $F_{PD,{E_n}}^{ipSIC}\left( x \right)$ of ${\gamma _{{E_n}}}$ for PD-NOMA with $K=1$ can be expressed as
\begin{align}\label{Corollary:the CDF of Eves for PD-NOMA}
&F_{PD,{E_n}}^{ipSIC}\left( x \right) = {\rm{Pr}}\left[ {\mathop {\max }\limits_{e \in {\Phi _e}} \left( {\frac{{{\rho _e}{{\left| {{{\tilde h}_{ek}}} \right|}^2}{a_n}}}{{\varpi {\rho _e}{{\left| {{h_{{I_e}k}}} \right|}^2} + 1}}} \right) < x} \right]  \nonumber\\
\
&= {{\mathop{\rm \mathbb{E}}\nolimits} _{{\Phi _e}}}\left\{ {\prod\limits_{e \in {\Phi _e}} {\int_0^\infty  {{f_{\tilde Z}}\left( z \right){F_{\tilde Y}}\left( {\frac{{x\left( {{\rho _e}z + 1} \right)\left( {1 + d_e^\alpha } \right)}}{{{\varphi _1}}}} \right)} dy} } \right\},
\end{align}
where $\varpi \ne 0$.

Applying the generating function, we can re-write \eqref{Corollary:the CDF of Eves for PD-NOMA} as
\begin{align}\label{Corollary:the CDF of Eves for PD-NOMA 1}
&F_{PD,{E_n}}^{ipSIC}\left( x \right) = \exp \left\{ { - 2\pi {\lambda _e}\int_0^\infty  {\left[ {1 - \int_0^\infty  {{f_{\tilde Z}}\left( z \right)} } \right.} } \right. \nonumber \\
  &\times \left. {\left. {{F_{\tilde Y}}\left( {\frac{{x\left( {{\rho _e}z + 1} \right)\left( {1 + {r^\alpha }} \right)}}{{{\varphi _1}}}} \right)dy} \right]rdr} \right\} \nonumber \\
    &= \exp \left\{ { - 2\pi {\lambda _e}{\varphi _1}\underbrace {\int_0^\infty  {\frac{{r{e^{ - \frac{{x\left( {1 + {r^\alpha }} \right)}}{{{\varphi _1}}}}}}}{{{\varphi _1} + x{\rho _e}{\Omega _{{I_e}}} + x{\rho _e}{\Omega _{{I_e}}}{r^\alpha }}}} dr}_{{J_4}}} \right\}.
\end{align}

By invoking \cite[Eq. (3.383.10)]{gradshteyn}, we can obtain
\begin{align}\label{J4}
{J_4} = &\Gamma \left( \delta  \right){e^{\frac{1}{{{\rho _e}{\Omega _{{I_e}}}}}}}\frac{{{{\left( {{\varphi _1} + x{\rho _e}{\Omega _{{I_e}}}} \right)}^{\delta  - 1}}}}{{\alpha {{\left( {x{\rho _e}{\Omega _{{I_e}}}} \right)}^\delta }}}  \nonumber \\
    & \times \Gamma \left( {1 - \delta ,\frac{{{\varphi _1} + x{\rho _e}{\Omega _{{I_e}}}}}{{{\rho _e}{\Omega _{{I_e}}}{\varphi _1}}}} \right).
\end{align}

Upon substituting \eqref{J4} into \eqref{Corollary:the CDF of Eves for PD-NOMA 1}, the CDF of $F_{PD,{E_n}}^{ipSIC}\left( x \right)$ can be given by
\begin{align*}\label{Corollary:the CDF of Eves for PD-NOMA 2}
& F_{PD,{E_n}}^{ipSIC}\left( x \right) = \exp \left\{ { - \delta \pi {\lambda _e}{\varphi _1}\Gamma \left( \delta  \right){e^{\frac{1}{{{\rho _e}{\Omega _{{I_e}}}}}}}} \right. \nonumber \\
\end{align*}
\begin{align}
 & \left. { \times \frac{{{{\left( {{\varphi _1} + x{\rho _e}{\Omega _{{I_e}}}} \right)}^{\delta  - 1}}}}{{{{\left( {x{\rho _e}{\Omega _{{I_e}}}} \right)}^\delta }}}\Gamma \left( {1 - \delta ,\frac{{{\varphi _1} + x{\rho _e}{\Omega _{{I_e}}}}}{{{\rho _e}{\Omega _{{I_e}}}{\varphi _1}}}} \right)} \right\},
\end{align}
where $\delta  = \frac{2}{\alpha }$.

With the help of derivation formula, i.e., ${\Gamma \left( {s,x} \right)}|_x^{'} = - {x^{s - 1}}{e^{ - x}}$ and deriving the above formula, the PDF of $f_{PD,{E_n}}^{ipSIC}\left( x \right)$ can be obtain
\begin{align}\label{Corollary:the PDF of Eves for PD-NOMA}
&f_{PD,{E_n}}^{ipSIC}\left( x \right) = {e^{\Theta \frac{{{{\left( {{\varphi _1} + x{\varphi _2}} \right)}^{\delta  - 1}}}}{{{{\left( {x{\varphi _2}} \right)}^\delta }}}\Gamma \left( {1 - \delta ,\frac{{{\varphi _1} + x{\varphi _2}}}{{{\varphi _1}{\varphi _2}}}} \right)}}\Theta  \nonumber \\
  &\times \left\{ {\left[ {\frac{{\left( {\delta  - 1} \right){{\left( {{\varphi _1} + x{\varphi _2}} \right)}^{\delta  - 2}}}}{{{x^\delta }\varphi _2^{\delta  - 1}}} - \frac{{\delta {{\left( {{\varphi _1} + x{\rho _e}{\Omega _{{I_e}}}} \right)}^{\delta  - 1}}}}{{{x^{\delta  + 1}}\varphi _2^\delta }}} \right]} \right. \nonumber\\
 & \left. { \times \Gamma \left( {1 - \delta ,\frac{{{\varphi _1} + x{\varphi _2}}}{{{\varphi _1}{\varphi _2}}}} \right) - \frac{{\varphi _1^{\delta  - 1}{e^{ - \frac{{{\varphi _1} + x{\varphi _2}}}{{{\varphi _1}{\varphi _2}}}}}}}{{{x^\delta }\left( {{\varphi _1} + x{\varphi _2}} \right)}}} \right\},
\end{align}
where $\Theta  =  - \delta \pi {\lambda _e}{\varphi _1}\Gamma \left( \delta  \right){e^{\varphi _2^{ - 1}}}$.

Upon substituting \eqref{Corollary:the PDF of Eves for PD-NOMA} and \eqref{PD-NOMA:the CDF of SINR expression for the near LUs with ipSIC} into \eqref{CD-NOMA:the expression of COP for near user}, we can obtain \eqref{the expression for the n-th user with ipSIC for PD-NOMA} and complete the proof.

\appendices
\section*{Appendix~D: Proof of Theorem  \ref{Theorem:the SOP of the m-th user for CD-NOMA}} \label{Appendix:D}
\renewcommand{\theequation}{D.\arabic{equation}}
\setcounter{equation}{0}

The proof starts by solving the CDF $F_{{E_m}}^{CD}\left( {\rm{x}} \right)$ of ${\gamma _{{E_m}}}$ for CD-NOMA. Based on \eqref{the SINR expression of eavesdropper to detect the m-th user}, we can formulate
\begin{align}\label{Appenxix D: The CDF of Eves to wiretap the m-th user for CD-NOMA}
F_{{E_m}}^{CD}\left( {\rm{x}} \right) &= {\rm{Pr}}\left[ {\mathop {\max }\limits_{e \in {\Phi _e}} \left( {\frac{{{\rho _e}{a_m}\left\| {diag\left( {{{\bf{h}}_e}} \right){{\bf{g}}_e}} \right\|_2^2}}{{{\rho _e}{a_n}\left\| {diag\left( {{{\bf{h}}_e}} \right){{\bf{g}}_e}} \right\|_2^2 + 1}}} \right) < x} \right]\nonumber \\
& = {\mathop{\rm \mathbb{E}}\nolimits} \left\{ {\prod\limits_{e \in {\Phi _e}} {{F_Y}\left[ {\frac{{\left( {1 + d_e^\alpha } \right)x}}{{\eta {\rho _e}\left( {{a_m} - {a_n}x} \right)}}} \right]} } \right\},
\end{align}
where ${a_m} > x{a_n}$ and the CDF of ${F_Y}\left( y \right)$ can be obtain from Appendix B.

Following a procedure similar to that used for obtaining \eqref{the CDF of EE with ipSIC for CD-NOMA}, we employ the generating function and turn to polar coordinates. Then \eqref{Appenxix D: The CDF of Eves to wiretap the m-th user for CD-NOMA} can be further expressed as
\begin{align}\label{Appenxix D: The derive process of CDF Eves to wiretap the m-th user for CD-NOMA}
& F_{{E_m}}^{CD}\left( {\rm{x}} \right) = \exp \left\{ { - 2\pi {\lambda _e}{e^{\frac{{ - x}}{{\eta {\rho _e}\left( {{a_m} - {a_n}x} \right)}}}}} \right.\sum\limits_{i = 0}^{K - 1} {\frac{1}{{i!}}}  \nonumber \\
 & \left. { \times {{\left[ {\frac{x}{{\eta {\rho _e}\left( {{a_m} - {a_n}x} \right)}}} \right]}^i}\int_0^\infty  {{{\left( {1 + {r^\alpha }} \right)}^i}{e^{\frac{{ - x{r^\alpha }}}{{\eta {\rho _e}\left( {{a_m} - {a_n}x} \right)}}}}rdr} } \right\}  .
\end{align}
Applying Binomial theorem and \cite[Eq. (3.326.2)]{gradshteyn}, we arrive at
\begin{align}\label{Appenxix D: The derive process of CDF Eves to wiretap the m-th user for CD-NOMA}
&F_{{E_m}}^{CD}\left( x \right)= \exp \left\{ { - \delta \pi {\lambda _e}\sum\limits_{i = 0}^{K - 1} {\sum\limits_{j = 0}^i {{
   i  \choose
   j  }} } \frac{1}{{i!}}} \right.  \nonumber \\
&\left. { \times \Gamma \left( {j + \delta } \right){{\left[ {\frac{{\eta {\rho _e}\left( {{a_m} - {a_n}x} \right)}}{x}} \right]}^{j - i + \delta }}{e^{\frac{{ - x}}{{\eta {\rho _e}\left( {{a_m} - {a_n}x} \right)}}}}} \right\}.
\end{align}
Upon setting the derivation of the CDF in \eqref{Appenxix D: The derive process of CDF Eves to wiretap the m-th user for CD-NOMA}, we can obtain the PDF of
\begin{align}\label{Appenxix D: The derive process of CDF Eves to wiretap the m-th user for CD-NOMA}
&f_{{E_m}}^{CD}\left( x \right) = {e^{ - \Upsilon \sum\limits_{i = 0}^{K - 1} {\sum\limits_{j = 0}^i {{
   i  \choose
   j
}\frac{{\Gamma \left( {j + \delta } \right)}}{{i!}}} } {{\left[ {\frac{{\eta {\rho _e}\left( {{a_m} - {a_n}x} \right)}}{x}} \right]}^{j + \delta  - i}}}}\Upsilon
\nonumber \\
& \times \sum\limits_{i = 0}^{K - 1} {\sum\limits_{j = 0}^i {{
   i  \choose
   j
}} \frac{{\Gamma \left( {j + \delta } \right){{\left( {\eta {\rho _e}} \right)}^{j + \delta  - i}}}}{{i!}}} \left[ {\frac{{{a_n}\left( {j + \delta  - i} \right)}}{{{x^{j + \delta  - i}}\left( {{a_m} - {a_n}x} \right)}}} \right. \nonumber\\
  & \left. { + \frac{{{a_m}}}{{\eta {\rho _e}{x^{j + \delta  - i}}{{\left( {{a_m} - {a_n}x} \right)}^2}}} + \frac{{\left( {j + \delta  - i} \right)}}{{{x^{j + \delta  - i{\rm{ + }}1}}}}} \right]{\left( {{a_m} - {a_n}x} \right)^{j + \delta  - i}},
\end{align}
where $\Upsilon   = \delta \pi {\lambda _e}{e^{\frac{{ - x}}{{\eta {\rho _e}\left( {{a_m} - {a_n}x} \right)}}}}$.

By the virtue of \cite[Eq. (7)]{Yue8370069Unified} and removing the ordering operation,
the CDF $F_{{\gamma _m}}^{CD}$ of the $m$-th user for CD-NOMA can be given by
\begin{align}\label{Appenxix D: the CDF of far user for CD-NOMA}
F_{{\gamma _m}}^{CD}\left( x \right) \approx &\sum\limits_{u = 1}^U {{b_u}} \left[ {1 - {e^{ - \frac{{x{c_u}}}{{\eta \rho \left( {{a_m} - x{a_n}} \right)}}}}} \right. \nonumber\\
 & \times \left. {\sum\limits_{i = 0}^{K - 1} {\frac{1}{{i!}}{{\left( {\frac{{x{c_u}}}{{\eta \rho \left( {{a_m} - x{a_n}} \right)}}} \right)}^i}} } \right] ,
\end{align}
where ${a_m} > x{a_n}$.

Substituting \eqref{Appenxix D: The derive process of CDF Eves to wiretap the m-th user for CD-NOMA} and \eqref{Appenxix D: the CDF of far user for CD-NOMA} into \eqref{the expression of SOP for the m-th user}, we can obtain \eqref{the SOP of the m-th user for CD-NOMA} and complete the proof.

\bibliographystyle{IEEEtran}
\bibliography{mybib}

\end{document}